\documentclass{jpp}
\usepackage{url}
\usepackage{ragged2e}
\usepackage{graphicx} 
\usepackage[table,xcdraw]{xcolor}
\usepackage{mathtools}
\usepackage{optidef}
\usepackage{subcaption}
\usepackage{cellspace} 
\newlength{\subcolumnwidth}
\newenvironment{subcolumns}[1][0.45\columnwidth]
 {\valign\bgroup\hsize=#1\setlength{\subcolumnwidth}{\hsize}\vfil##\vfil\cr}
 {\crcr\egroup}
\newcommand{\nextsubcolumn}[1][]{%
  \cr\noalign{\hfill}
  \if\relax\detokenize{#1}\relax\else\hsize=#1\setlength{\subcolumnwidth}{\hsize}\fi
}
\newcommand{\nextsubfigure}{\vfill}

\shorttitle{Designing A Buildable Optimized Stellarator for Pair Plasmas}
\shortauthor{P.F. Gil et al.}

\title{Designing A Buildable Optimized Stellarator to Confine Electron-Positron Plasmas}

\author{Pedro F. Gil\aff{1}
  \corresp{\email{pedro.gil@ipp.mpg.de}},
   Jason Smoniewski\aff{2,3}, Paul Huslage\aff{1}, Rogerio Jorge\aff{4}, Timo Thun\aff{2}, Elisa Buglione-Ceresa\aff{1}, Tristan Schuler \aff{5}, Stefan Fingl \aff{1}, Grégoire-Hubert Ducas\aff{6}, \and Eve V. Stenson\aff{1}}

\affiliation{\aff{1}Max-Planck-Institut für Plasmaphysik, 85748 Garching, Germany
\aff{2}Max-Planck Institute for Plasma Physics,
 17491 Greifswald, Germany
 \aff{3} Plasma Science and Fusion Center, Massachusetts Institute of Technology, Cambridge, Massachusetts 02139, USA
 \aff{4}Department of Physics, University of Wisconsin-Madison, Madison, Wisconsin 53706, USA
\aff{5}SchulerTEC, Germany
\aff{6}Department of Mechanical Engineering, McGill University, Montreal, QC, 13 H3A 0B9, Canada}

\begin{document}

\maketitle

\begin{abstract}
\justifying
In this paper, the design of the the plasma equilibrium and superconducting coils for the Electrons and Positrons in an Optimized Stellarator EPOS experiment is presented. With newly developed stellarator optimization tools, including single-stage and stochastic optimization, as well as HTS strain, this work demonstrates that it is possible to achieve key metrics for the buildability and confinement properties of the device. In particular, satisfactory quality of quasisymmetry and stellarator robustness is designed, and engineering requirements are met for eight different candidates. A feasibility study is presented that optimizes multiple candidates for different plasma major radii and coil currents, as well as the best EPOS candidate to date.
\end{abstract}

\section{Introduction}
\justifying

In magnetically confined fusion, deuterium and tritium are heated in order to produce an electron-ion plasma that is kept inside a magnetic field. This, however, has been revealed to be a highly complicated task, partly due to neoclassical and turbulent transport. One of the key drivers of a series of instabilities found in these plasmas is mass asymmetry, as electrons and ions react on different timescales. In electron-positron pair plasmas, in contrast, it is predicted that turbulent transport is suppressed, meaning that the behavior is quiescent compared to its fusion counterpart \citep{tsytovich,Helander_microstrability,Helander_Connor_2016}. There has been little experimental exploration of pair plasmas with electrons and positrons. Developing new devices capable of confining these entities could provide further insights into astrophysical plasmas, such as those found near pulsars and neutron stars \citep{Arons1979SomePO}. 
Designing an experiment suitable for reproducing an electron-positron plasma on Earth requires the conception of a magnetic field configuration that can confine the two species stably for a sufficiently long time. The stellarator configuration is a suitable candidate, as it is an inherently steady-state machine,  and it does not require driving currents in the plasma. External 3D-shaped coils fully determine the shape of the magnetic field. Recent work \citep{LandremanAndPaul} has shown that it is possible to optimize stellarator equilibria for precise quasisymmetry across all the magnetic surfaces. Quasisymmetry is a measure of how well collisionless trapped particles are confined. Generating enough positrons to meet the density requirements for a pair plasma is challenging, meaning that any loss of positrons is to be avoided; by being quasisymmetric, EPOS is able to confine the particles for longer which in turn increases the effects of cyclotron cooling. This is desired as cooling down the particles by radiating the perpendicular energy allows for longer confinement times. This goes against typical fusion-related plasma experiments that want to heat the plasma. Assuming an optimized stellarator, this would mean that limitations on confinement time originate from collisionless transport driven by real-world device imperfections and/or from collisional or turbulent transport.
\\

For the EPOS stellarator, an on-axis field of 2\,T is targeted to achieve cooling of the plasma through cyclotron radiation within less than a second \citep{Kennedy_Helander_2021}. This makes ReBCO (Rare earth Barium Copper Oxide) HTS (high-temperature superconductors) a promising candidate to realize the EPOS magnet system. ReBCO superconductors have a high in-field performance and a critical temperature which can be comfortably achieved with compact cryocoolers without using cryogenic liquids. Non-insulated (NI) coils made from ReBCO superconductors show high thermal and electrical stability \citep{hahn_hts_2011}. Non-planar NI coils have been demonstrated at low fields \citep{huslage_winding_2024, kim_numerical_2025}. We aim to use this approach for the EPOS coils in same spirit as described in \cite{Baillod_2025_1, Baillod_2026_1}. 


Another aspect that has seen significant progress is the specific field of stellarator coil optimization. Many codes have been proposed to optimize and find 3D shaped coils, such as COILOPT \citep{strickler}, FOCUS \citep{Zhu_2018,HUDSON20182732}, or REGCOIL \citep{regcoil}. Initially, these codes employed a two-stage approach: the first stage optimized the equilibrium to meet plasma specifications, and the second stage shaped the coils to closely reproduce the magnetic field generated in the first step. This approach can lead to incompatibility problems where the equilibria are far too complex to be reproduced by any buildable coils. To circumvent these issues, recent work by \cite{Henneberg_2021,Jorge_2023,Giuliani_Wechsung_Stadler_Cerfon_Landreman_2022, giuliani2} has enabled us to combine the two-step approach into a single-stage stellarator optimization.

The paper is structured as follows: Section \ref{sec:epos_requirements} presents both the physics and engineering requirements for EPOS, outlining the targets that guide the optimization. Then, in Section \ref{sec:optimization_method}, the techniques used for the coils and plasma optimization in order to achieve the various configurations within the physics and engineering requirements are presented. Section \ref{sec:results} presents the results obtained from eight candidate configurations and provides information on the decision-making process leading to a final configuration. Finally, in Section \ref{sec:conclusion}, the results are summarized, and ideas for future improvement and analysis of the EPOS device are presented as the project advances towards its assembly and manufacturing phase.

\section{EPOS Requirements}
In this section, we explain how EPOS differs from other fusion-targeted stellarators and how this affects the optimization of physics and engineering objectives. Size and magnetic field strength are two main drivers for the unique design of EPOS.
\label{sec:epos_requirements}
\subsection{\textit{Physics Requirements}}
As mentioned previously, the goal of EPOS is to confine electrons and positrons in a stellarator and to observe collective effects from the plasma. EPOS is hence different from its fusion counterparts. It is expected that EPOS will confine $10^{10}-10^{11}$ positrons and electrons, which means that it is going to be a low $\beta$ plasma ($\beta <10^{11}$\%). In order to achieve conditions suitable for observable collective effects, a sufficiently high $a / \lambda_D$ ratio is required, where $a$ is the device's minor radius and $\lambda_D$ is the Debye length, such a ratio measures how large the device is in comparison to the characteristic shielding length scale, which scales as:

\begin{equation}
    \frac{a}{\lambda_D} = a \sqrt{\frac{2 n_p e^2}{\epsilon_0 k_B T_p}} \sim \sqrt{\frac{N_p}{T_p R}}.
\label{eq:debye_length}
\end{equation}
Here, R corresponds to the machine's major radius, $N_p$, $n_p$, $T_p$, and e refer to the number, density, temperature, and charge of the positrons, respectively. EPOS therefore requires minimizing the major radius while maximizing the magnetic field amplitude in order to reach low enough temperatures.
EPOS will rely on cyclotron radiation cooling of the particles to a range of about 0.1 - 1 eV; this means that the magnetic field on axis should reach $\approx$ 2 T. For cyclotron radiation cooling to be efficient, confinement times on the order of 1 s are necessary \citep{pedersen2012}. In addition to the two design choices mentioned previously, namely machine size and magnetic field amplitude, which directly affect the optimization scheme, the existence of the so-called "weave-lane" (WL) coils is also a consideration. The latter will be explained in the next section. \\

Firstly, regarding dimensions, early estimations of device size show that a desirable upper boundary on the plasma volume is of ~10 L \citep{pedersen2012}. This translates to a first approximation of a major radius of 20 cm and a minor radius of 5 cm. The use of these values in the optimization will be explained in Section \ref{sec:optimization_method}. 
To fulfill the size and field requirements, High-Temperature Superconductors (HTS) are suitable candidates for the coils. However, this brings geometrical consequences, meaning that the coils are not allowed to have prohibitive binormal curvature or torsion in order not to compromise the brittle nature of the HTS. Finally, considering the choice of a quasisymmetric field to confine collisionless particles for a sufficiently long time to cool them down, the quasi-axisymmetric (QA) configuration appears to be the most suitable for fulfilling EPOS requirements, namely high field strengths at small dimensions, thereby avoiding high coil complexity. This can be achieved because QA configurations are notably simpler than their quasi-helically symmetric and quasi-isodynamic relatives, as they strongly resemble a tokamak. 

\begin{table} 
\centering
\resizebox{6cm}{!}{
\renewcommand{\arraystretch}{2}
\begin{tabular}{|l|c|c|}
\hline
\rowcolor[HTML]{C0C0C0} 
\textbf{EPOS Parameters} & \multicolumn{1}{l|}{\cellcolor[HTML]{C0C0C0}\textbf{Value / Type}} & \multicolumn{1}{l|}{\cellcolor[HTML]{C0C0C0}\textbf{Unit}} \\ \hline
$\vert B\vert$ on axis              & 2           & T                             \\ \hline
Particle Energy          & 0.1-1                                                              & eV                                                         \\ \hline
a / $\lambda_D$            & $>$ 10                                                               & -                                                          \\ \hline
Density                  & $10^{10}$ - $10^{11}$                                                      & $m^{-3}$                                                    \\ \hline
Plasma Volume            & 10                                                                 & L                                                          \\ \hline
\end{tabular}
}
\caption{Summary of the main physics parameters of EPOS.}
\label{tab:epos_parameters}
\end{table}

The choice of a QA field together with HTS coils limits the magnitude of the rotational transform $\iota$ that is required for EPOS. Unlike fusion stellarators, where the profile of the rotational transform can have a role controlling effects such as the Shafranov-Shift, ballooning modes, and Mercier stability \citep{Hegna}, for EPOS, a flat rotational transform profile is desired. By keeping its value constant radially, it is possible to avoid hitting any low-order resonant modes that could give rise to magnetic islands breaking up the flux surfaces and damaging the confinement quality. Moreover, early optimization attempts revealed that the upper boundary of iota is determined by the trade-off between the HTS strain targeted to remain below the threshold and a low quasisymmetric error. Table \ref{tab:epos_parameters} summarizes the primary physics targets for EPOS.

\subsection{\textit{Engineering Requirements}}
\begin{figure} 
    \centering
    \includegraphics[scale = 0.19]{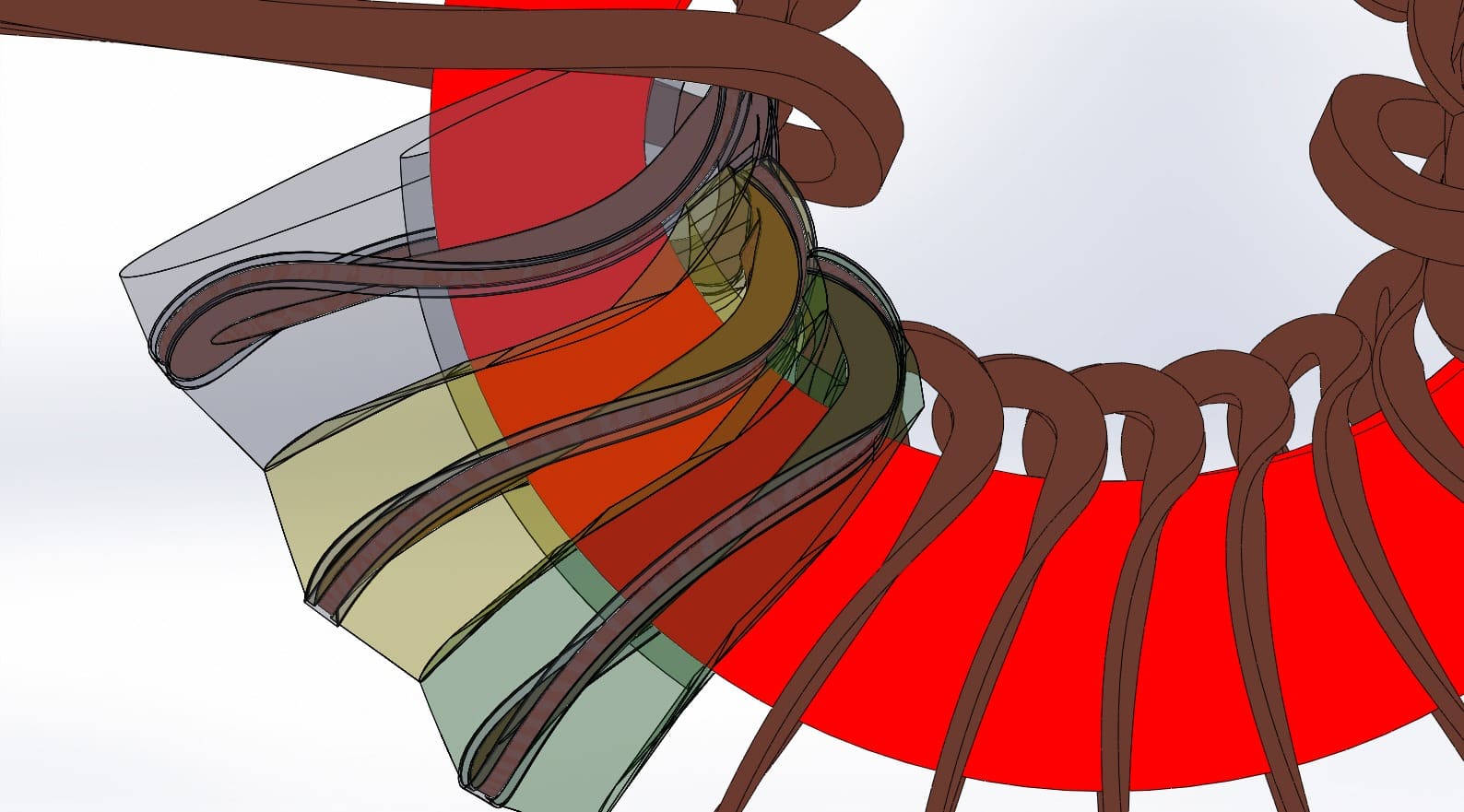}
    \caption{Top view of the engineering design concept for EPOS coils, each coil is made of a double HTS winding pack (here shown as a single brown stack) and is supported in an individual stellarator frame shown here in different colors for three coils. The last closed flux surface is shown in red.}
    \label{fig:stellarator_slice}
\end{figure}

In this section, the main engineering challenges are outlined when assessing the feasibility of a machine of the same dimensions as EPOS. As mentioned in the previous section, HTS will be the material of choice for the coils, and keeping its mechanical integrity via the minimization of the binormal curvature and torsion is of high priority. For that matter, as mentioned in the previous section, it is essential to keep the strain of the winding pack for both torsion and binormal curvature below a threshold of 0.2 $\%$ \citep{huslage2024strainoptimizationrebcohightemperature}. 
The HTS coils will consist of double pancakes, meaning two adjacent HTS tape stacks, which will be wound around a 3D metal frame. At this stage, it is planned to mill the trenches in which the tapes will lie using a 5-axis CNC machine. Each coil will be demountable and have its own unit, as shown in Figure \ref{fig:stellarator_slice}. Each double pancake possesses a 1 mm wall in the middle. A single pancake is 3 mm wide and approximately 17 mm tall, assuming a 0.1 mm spacing between HTS layers for solder paste and a tape height of 0.1 mm. The WL coils will consist of either 2, 3, or 4 double pancakes. The trenches will have some additional slack in width to accommodate winding the coil; early estimates indicate that 0.5 mm of slack should be sufficient for comfortable winding.

The milling and winding processes introduce several inaccuracies that accumulate, resulting in an overall imprecision of the current path geometry, commonly referred to as coil perturbations. These deviations are a notable source of problems in modern stellarators due to manufacturing tolerances, such as W7-X \citep{w7x_perturbations} and NCSX \citep{ncsx}, as they can potentially lead to poor confinement, the formation of magnetic islands, and possibly the destruction of flux surfaces. As a reference, W7-X was built with a magnetic field amplitude error tolerance of $10^{-4}$ at 2 mm coil deviation amplitude ($\delta_{W7-X}$). Assuming the scaling is linear, this would lead to the following tolerances for EPOS:

\begin{equation}
    \delta_{EPOS} = \frac{\delta_{W7-X}a_{EPOS}}{a_{W7-X}} = \frac{1e-3 \times 0.1}{1} \approx 10^{-4} \space \text{m}
\end{equation}

Tolerances of $10^{-4}$ m, although not currently achievable with metal 3D printing technology, can be met via milling, which operates with sub-$10^{-4}$ m precision. However, the extra slack mentioned previously, present in the width of the trenches, means that EPOS should still be able to present a precise enough QA field with a maximum deviation of 0.5 mm from trench uncertainty. Therefore, it is crucial to ensure that EPOS meets these robustness requirements, and one way to do so is through stochastic optimization, which is described in Section \ref{sec:stoch_section}.
\\

Another aspect that has to be taken into account when optimizing EPOS are the $\textbf{J}\times \textbf{B}$ forces. This can be of particular concern on the inboard side of the stellarator, where the coils find themselves closer to each other; an estimate for the Lorentz forces is shown in the Appendix \ref{sec:forces}. These forces are induced by the other coils and the coil itself. If not considered, they can sometimes reach magnitudes high enough to deform metal structures, particularly in fusion devices. As a conservative measure EPOS will consist of a segmented shell, and it is on this shell that the trenches containing the HTS stacks take shape in several coils frames. As a result, EPOS will not have singular isolated modular coils, and the shapes of the coils will also ensure that the trenches remain accessible for winding. The coil frames are designed to allow for optimal alignment (minimizing assembly errors) and conductive cooling of the HTS tapes.

\begin{figure} 
    \centering
    \includegraphics[width=0.45\linewidth]{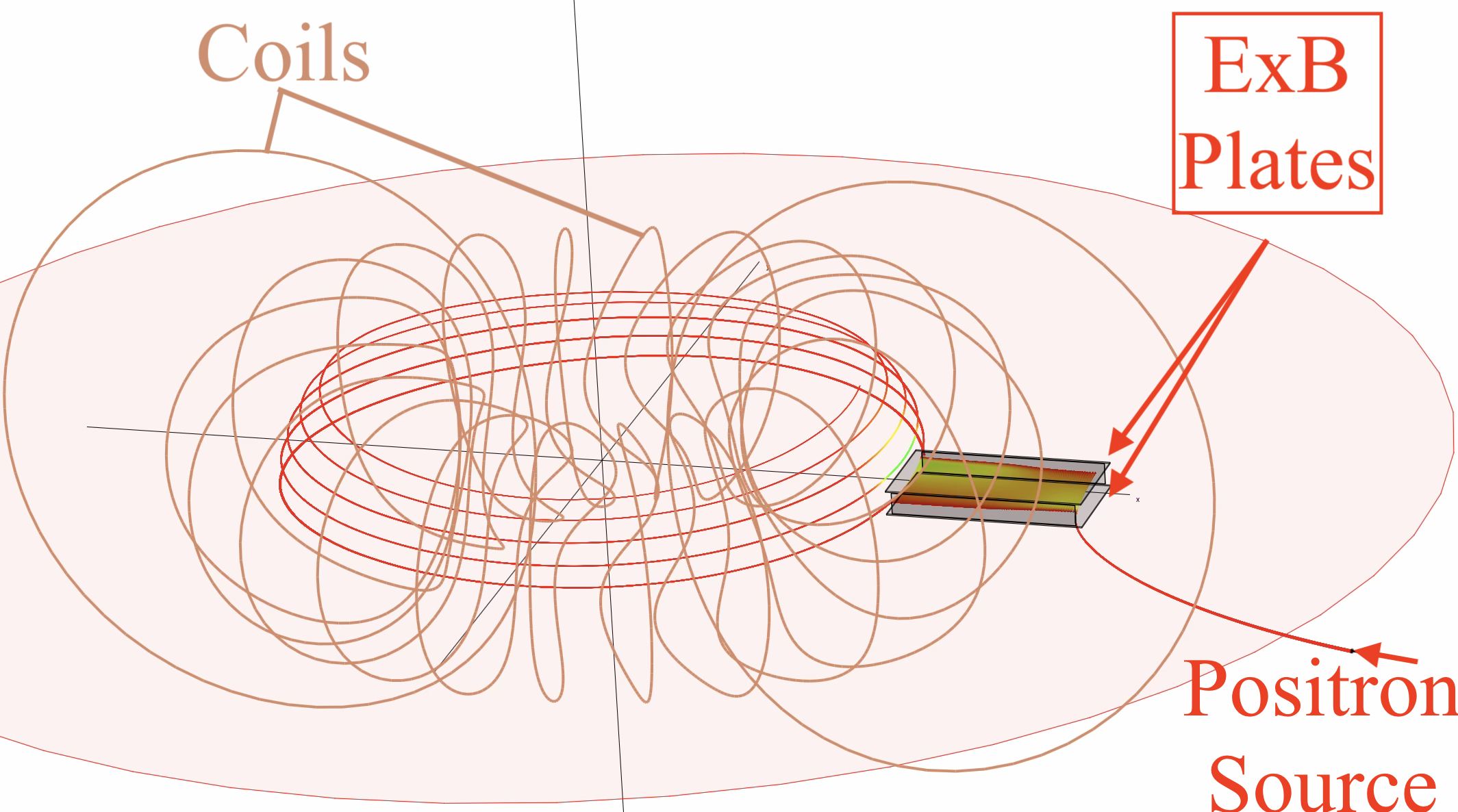}
    \includegraphics[width=0.45\linewidth]{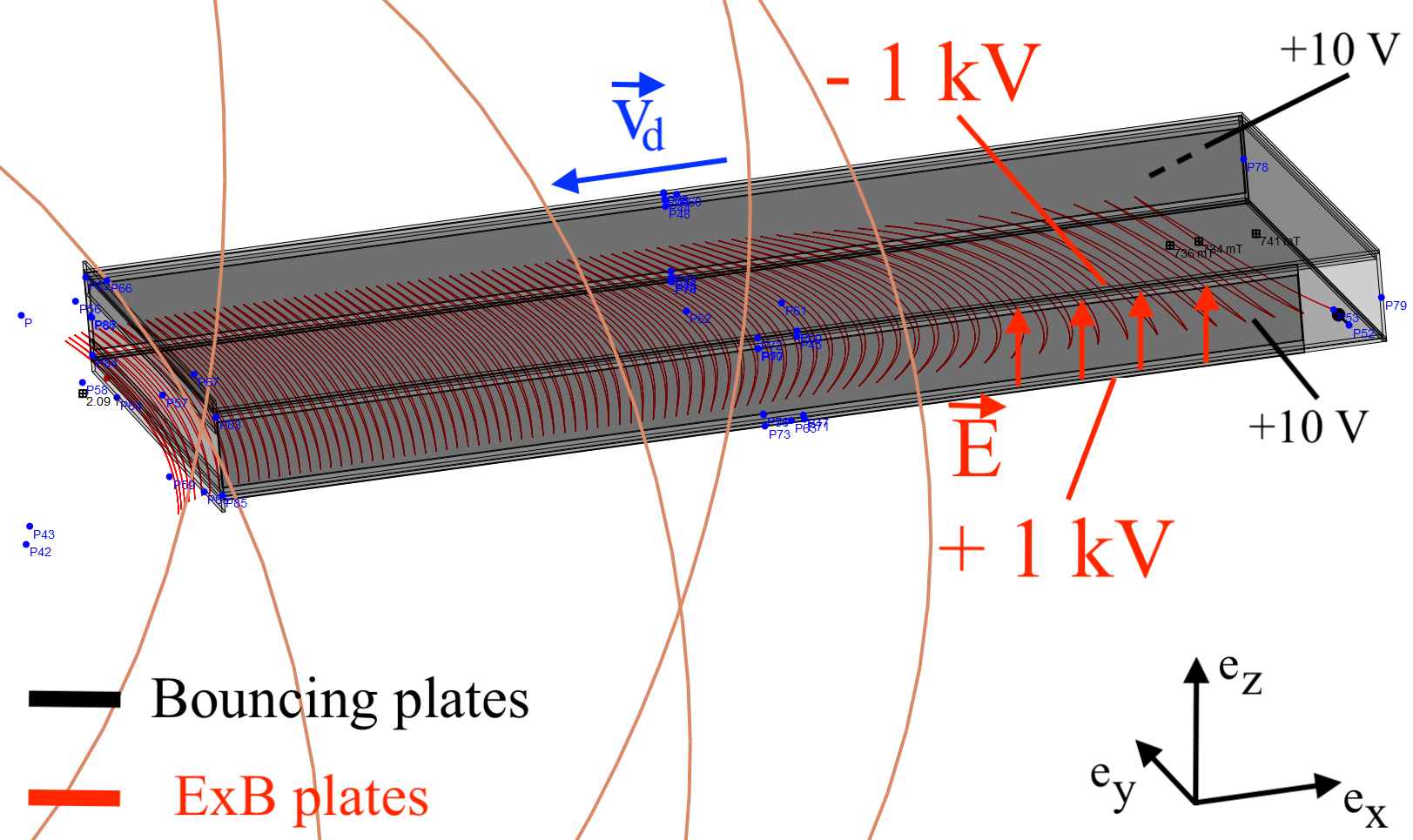}
    \caption{\textbf{Left}: Injection concept for the EPOS stellarator, a single positron trajectory is plotted in red starting at a field strength of about 6 mT and then channeled by a stray field line originating from the weave-lane coil. \textbf{Right}: the particles arrive at the E$\times$B electrodes and experience a drift in the radial direction as well as multiple bounces (from additional bouncing plates) until it is finally injected into the 3D shaped field. The electrons are to  be injected via an emitter.}
    \label{fig:ExB_injection}
\end{figure}

Finally, the previously mentioned "weave-lane" coils are required for the injection of positrons via $\textbf{E}\times\textbf{B}$ drift into the stellarator, similar to the injection technique into a magnetic dipole described in \cite{dipole_injection}. These consist of two additional coils, also made of HTS, that are placed in toroidally opposed positions and possess a radius and current that are large enough to generate stray field lines not connected to the interior of the device, hence the name "weave-lane". These field lines channel the positrons into two facing electrodes, inducing an electric field normal to the stray magnetic field in order to force the positrons into the magnetic field containing the flux surfaces, as shown in Figure \ref{fig:ExB_injection}. Additionally, two bouncing plates force the positrons into a back-and-forth motion as drift radially inwards. Since they will be considerably larger than the standard coils, maintaining a magnetic field strength of 2 T on axis requires a larger current. At this stage of the design process, it is estimated that the standard coils (SC) will possess approximately 70 kA, and the weave-lanes will carry even higher currents.

\section{Optimization Method}
\label{sec:optimization_method}
Here, the goal is to present how we can integrate the physics and engineering requirements explained in the previous section into state-of-the-art stellarator optimization and steer it towards an EPOS buildable configuration.
\subsection{\textit{Plasma Objective Function}}

To address the challenge of designing EPOS, the task relies heavily on computational optimization via the \texttt{SIMSOPT} framework \citep{landreman_simsopt_2021}. It offers a broad set of tools to analyze, build, and optimize stellarators. In our case, the challenge is twofold: find coils and an equilibrium that satisfy the aforementioned requirements. Regarding the physics part, \texttt{SIMSOPT} presents a Python interface with the Fortran Variational Moments Equilibrium Code (VMEC) \citep{vmec}, which minimizes the MHD potential energy following a variational principle approach by assuming the existence of flux surfaces and so contains information on the plasma equilibrium, close to the STELLOPT optimization suite \citep{stellopt}: 

\begin{equation}
    W = \int \left(\frac{p}{\gamma -1} + \frac{\vert B \vert^2 }{2\mu_0} \right)dV,
\label{eq:VMEC}
\end{equation}

where $W$ is the total MHD energy, $B$ is the magnetic field, $p$ is the plasma pressure, and $\gamma$ is the adiabatic index. It can be shown that a state that minimizes Equation \ref{eq:VMEC} verifies the force balance equation:

\begin{equation}
    J \times B = \nabla p.
\end{equation}
Here, $J$ is the plasma current density. Although EPOS will not reach a regime where MHD is relevant, the VMEC code also allows for solving vacuum fields, and more importantly, finding configurations with good quasisymmetry. In this work, it is operated in fixed-boundary mode, meaning that the last closed flux surface (LCFS) is fixed, serves as a boundary condition, and Equation \ref{eq:VMEC} within its volume. The LCFS is described through a Fourier decomposition:

\begin{subequations}
\begin{eqnarray}
    R(\theta, \phi) = \sum_{m=0}^{M_{pol}}\sum_{n=-N_{tor}}^{N_{tor}} \left[ a_{m,n}\cos(m\theta-n_{fp}n\phi) + b_{m,n}\sin(m\theta-n_{fp}n\phi) \right], \\
    Z(\theta, \phi) = \sum_{m=0}^{M_{pol}}\sum_{n=-N_{tor}}^{N_{tor}} \left[ c_{m,n}\cos(m\theta-n_{fp}n\phi) + d_{m,n}\sin(m\theta-n_{fp}n\phi) \right], 
\end{eqnarray}
 \label{eq:surface}
\end{subequations}
where $M_{pol}$ and $N_{pol}$ are the poloidal and toroidal modes respectively, and $n_{fp}$ is the number of field periods. The degrees of freedom that are optimized are the Fourier modes in the set of Equations \ref{eq:surface}: $x_S=(a_{m,n}, b_{m,n}, c_{m,n}, d_{m,n})$. It should be noted that when the magnetic field presents stellarator symmetry (i.e., $B(R,-\theta, -\phi) = B(R,\theta, \phi)$), the sine terms and cosine terms can be skipped for $R$ and $Z$, respectively. However, because of the weave-lane coils, EPOS does not have stellarator symmetry, which increases the number of surface degrees of freedom to 624. These degrees of freedom are then adjusted during the optimization to fulfill the following physics objective functions:

\begin{subequations}
    \begin{gather}
        f_{QS} = \sum_{s_i} w_i\biggr \langle \left[ \frac{1}{B^3}\left( (N-\iota M) \Vec{B} \times \nabla B \cdot \nabla \psi - (MG+NI) \Vec{B} \cdot \nabla B \right) \right]^2\biggr \rangle, \\
        f_{\iota} = (\iota - \iota_{t})^2, \\
        f_{A} = (A - A_{t})^2, \\
        f_{\Delta} = (\Delta-\Delta_t)^2, \\
        f_{S} = \left(\biggr \langle \frac{d\iota}{ds} \biggr \rangle - S_T\right)^2. 
    \end{gather}
\end{subequations}

Here, $f_{QS}$ is a function penalizing the departure from quasisymmetry, following the two-term formulation \citep{Rodríguez_Paul_Bhattacharjee_2022}. It is a weighted sum across all specified flux surfaces $s_i$ (10 equally spaced in our case), with $w_i$ its weights, B the magnetic field amplitude, and (N, M) are parameters specifying the type of quasisymmetry aimed at, with (1,0) corresponding to quasi-axisymmetry. $\psi$ is the toroidal flux function, and G is the poloidal and I the toroidal currents inside the surface. $f_{\iota}$ minimizes the difference between the surface's rotational transform and the target set by the user. $f_A$ and $f_w$ perform a similar task but regarding the aspect ratio A ( = R/a ) and the equilibrium radial width $\Delta$ of the bean cross-section. The latter is to prevent equilibria from violating the requirement of high $a/\lambda_D$. Finally, $f_S$ is used to minimize the magnetic shear, giving a flat iota radial profile.

\subsection{\textit{Coils Objective Function}}

For the coils, a method similar to the FOCUS code is presented, where coils are represented by geometric 3D curves in space and their shapes are varied during the derivative-based optimization. Similar to the method presented for the plasma surface, coil number i is represented by a Fourier decomposition of the form:

\begin{equation}
x^i(\theta)=\sum_{m=0}^{M_{coil}}x_{c,m}^i\cos(m\theta)+\sum_{n=0}^{M_{coil}}x_{s,n}^i\sin(n\theta),
\label{eq:coils_fourier}
\end{equation}

and same for $y^i$ and $z^i$. Note that $\theta$ here represents a coil length index and not the poloidal angle of the plasma. The order of decomposition is defined by the user. Equivalent to the plasma objective function, the coils degrees of freedom are given by the Fourier modes of Equation \ref{eq:coils_fourier}: $x_C=(x_{c,m}^i, x_{s,m}^i, z_{c,m}^i, z_{s,m}^i)$ $\forall i \in [1, N_{coils}]$ and $\forall m \in [0, M_{order}]$. EPOS is planned to have 11 coils per field period.

 The degrees of freedom of the coils present in Equation \ref{eq:coils_fourier} are also complemented with an additional decomposition of the winding angle of the tape. This is made in order to reduce the strain applied to tape when winding \citep{huslage2024strainoptimizationrebcohightemperature}. Using the same method defined in \citep{Singh_Kruger_Bader_Zhu_Hudson_Anderson_2020}, a centroid frame is constructed around the center coil filament using a modified Frenet-Serret frame. The winding angle is defined as the angle between the normal vector \textbf{$n_{centroid}$} in the centroid frame and the normal vector \textbf{n} of the actual HTS tape frame. This angle is then decomposed into a Fourier Series in the following way:

 \begin{equation}
    \alpha(l) = \alpha_0 + \sum_{n=1}^{N_{\alpha}}\alpha_{c,n}\cos(2\pi nl)+\alpha_{s,n}\sin(2\pi nl),
    \label{eq:angle_fourier}
 \end{equation}

where $N_{\alpha}$ is defined by the user and the coefficients $(\alpha_{c,n}, \alpha_{s,n})$ are additional degrees of freedom added to $x_C$. Note that $\alpha$ also coincides with the angle of the winding pack when considering the finite dimensions of the coils. Accounting for currents, $M_{order} = 7$ and $N_{\alpha}$, the total amount of coil degrees of freedom amounts to 575. 
Since coil optimization with 3D filaments is an ill-posed problem \citep{imbertgerard2020introductionstellaratorsmagneticfields}, additional regularization terms are added that allow for constraining the freedom of the optimizer when adapting the coil shapes. It is here that EPOS sets itself apart from conventional optimization for stellarator symmetric devices, as a differentiation for the weave-lane coils is necessary. Additional control over the shape, size, and current of the weave-lane coils is required, as they often prevent the optimizer from reaching good local minima. This is achieved by isolating specific functions in the optimization, enabling the setting of specific weights for the weave-lanes. The objective functions are therefore the following:

\begin{subequations}
    \begin{gather}
        f_{L}^{WL} =  \int_{\Gamma_0} dl, \\
        f_{L} = \sum_{i=2}^{N_{coils}} \int_{\Gamma_i} dl, \\
        f_{bin}^{WL} =  \frac{1}{2}\int_{\Gamma_0}\text{max}(\epsilon_{\text{bin}}^0 - \epsilon_{\text{bin}}^{max}, 0)^2 \, dl, \\
        f_{bin} =  \sum_{i=2}^{N_{coils}}\frac{1}{2}\int_{\Gamma_i}\text{max}(\epsilon_{\text{bin}}^i - \epsilon_{\text{bin}}^{max}, 0)^2 \, dl, \\
        f_{tor}^{WL} = \frac{1}{2}\int_{\Gamma_0}\text{max}(\epsilon_{\text{tor}}^0 - \epsilon_{\text{tor}}^{max}, 0)^2 \, dl ,\\
        f_{tor} = \sum_{i=2}^{N_{coils}}\frac{1}{2}\int_{\Gamma_i}\text{max}(\epsilon_{\text{tor}}^i - \epsilon_{\text{tor}}^{max}, 0)^2 \, dl, \\
        f_{msc}^{WL} = \frac{1}{L_0}\int_{\Gamma_0} \kappa^2\, dl, \\
        f_{msc} = \sum_{i=2}^{N_{coils}}\frac{1}{L_i}\int_{\Gamma_i} \kappa^2\, dl.
    \end{gather}
\end{subequations}

Additionally, $\Gamma$ is the coil path, and L is the total length of the coil, and $N_{coils}$ is the number of coils per period. Here, the superscript $f_{L}$ denotes the length of each coil, allowing for control over their size; $l$ represents the coordinate of the coil length. Then, $f_ {bin}$ targets the binormal curvature of the HTS tape used in the coils, where $\epsilon_{bin}^i$ is hard-way bending strain of coil number i, and $\epsilon_{tor}^{max}$ is equal to critical strain before the HTS tape breaks, in our case it is set to 0.2 \% conservatively when compared to the literature \citep{Soldan_2020} where a value of 0.4 \% is given. The same applies to $f_{tor}$ regarding torsional strain. Finally, the mean squared curvature $f_{msc}$ is also considered to prevent concave sections in the coils, which would significantly complicate the winding procedure of a single stack; here $\kappa$ represents the local curvature.

\subsection{\textit{Stochastic Objective Function }}
\label{sec:stoch_section}
To relax the tolerances on the coil shapes, an optimization approach was developed that utilizes stochastic coil optimization \citep{Wechsung_1,Wechsung_2,Lobsien_2018}. Coil stochastic optimization is specifically employed within the \texttt{SIMSOPT} framework. This method has been previously used for the design of the Columbia Neutral Torus, a stellarator comprising four coils: two interlocked and two poloidal field coils. There, stochasticity is considered by taking into account the tilts and translation errors of the coils. For EPOS, as described by Wechsung et al., the errors correspond to geometric deviations of the coil paths. The idea is to sample N random perturbed coils around the original coil and use them in the estimation of the field accuracy, measured through the quadratic flux metric, which will be explained in this section.

Considering the description of a coil as being a one dimensional current carrying filament of path $\gamma(l) = (X(l), Y(l), Z(l))$, with $l \in [0,1]$ the coordinate along the coil, a new small perturbation $\epsilon_i(l)$ modeled as a Gaussian process (GP) is added to each component of $\gamma$:

\begin{equation}
    \Tilde{\gamma_i}(l) = \gamma_i(l) + \epsilon_i(l).
\end{equation}
The GP can be interpreted as a generalization of random variables to functions, here the function is one of the cartesian coordinates $\gamma_i(l)$. It is described by a kernel $k(d)$, where $d=l_i-l_j$, which allows for the generation of a correlation function determining the regularity and smoothness of the perturbations along the coil. In this work, the RBF kernel is used, where:

\begin{equation}
    k(d) = \sigma^2 \exp{\left(\frac{-d^2}{2L_s^2}\right)},
\end{equation}
here $\sigma>0$ measures the amplitude of the perturbations and $L_s>0$ determines the characteristic length over which they occur. Moreover, in order to use this method in the optimization routine, it is necessary to extract the derivatives of the GP and add them to the covariance matrix $\Sigma$. Then, after performing a Cholesky decomposition such that $\Sigma = DD^\top$ and by drawing a vector $X = (X_1,..., X_n)$ from a multivariate normal distribution, where n is the number of quadrature points of the filaments, one vector of perturbation is generated by calculating $\epsilon_i = DX$. 

Following this, N samples corresponding to N different perturbed stellarators are drawn. Moreover, in the optimization method employed here, the shaping of the coils is driven by the quadratic flux metric. It is a metric for the precision with which the coils reproduce the intended magnetic field:

\begin{equation}
    f_{SF} = \frac{1}{2}\int_{S} \frac{\vert \Vec{B} \cdot \Vec{n}\vert^2}{\vert B \vert ^2}dS,
    \label{eq:squared_flux}
\end{equation}
Here is the LCFS, $\Vec{B}$ is the magnetic field generated by the coils, calculated through the Biot-Savart law, and $\Vec{n}$ is the local normal to the surface, calculated from the plasma object. This term bridges the two objects that were previously presented, the coils and the plasma.

Finally, from the methods described above, it is therefore possible to compute N different quadratic fluxes from the N different stellarators. In a sample average approximation, the average squared flux is estimated about the single unperturbed VMEC surface object:

\begin{equation}
    \langle f_{SF} \rangle = \frac{1}{N} \sum_{i=1}^N \frac{1}{2}\int_{S} \frac{\vert \Vec{B_i} \cdot \Vec{n}\vert^2}{\vert B_i \vert ^2}dS.
    \label{eq:squared_flux_stoch}
\end{equation}
The derivatives about the surface and coils degrees of freedom used in the gradient calculations \citep{Jorge_2023} are therefore also the corresponding averages. It is worth noting that the stochasticity here is only applied in the estimation of the squared flux and not in the regularization terms as they are not of relevance for the error calculations and could result in considerable discrepancies in the gradients amplitude (e.g. mean squared curvature estimation) and prevent good convergence.

\subsection{\textit{Combined Objective Function }}
Traditionally, stellarator design is performed in two steps: first, the plasma shape and properties are optimized, and then a second stage, where suitable coils are found that reproduce the previous magnetic field. However, this method has an inherent issue: the equilibrium found in the first step may not be reproduced to desirable accuracy, given the engineering constraints. In this regard, recent work by \citep{Jorge_2023} and tested in \cite{Baillod_2025_1} combines the two steps into a single-stage approach. This enables both the plasma and the coils to be optimized simultaneously, and grants a higher compatibility between the coils and the equilibrium; this is the method of choice used to optimize EPOS. At this point, it is possible to write the minimization problem as follows:

\begin{equation}
\begin{aligned}
\min_{x_C, x_S} \quad & F(x_C, x_S) = F_{S}(x_S) + w_{C}F_{C}(x_C, x_S).\\
\textrm{s.t.} \quad & \psi = \psi_0\\
  &R = R_0    \\
\end{aligned}
\label{eq:full_obj_ftion}
\end{equation}
The $w_C$ term is the weight that tunes the optimization towards or away from coil targets to the detriment of surface objectives. In Equation \ref{eq:full_obj_ftion}, the full objective function is a linear combination of the surface objective function $F_S$ that is determined by the LCFS degrees of freedom $x_S$ and the coils objective function $F_C$ that is dependent on both the surface's and the coils' degrees of freedom since the quadratic flux is included in it. An additional constraint is applied to the parameter space, as the major radius is kept fixed throughout the optimization. This also comes at a cost for the optimizer, which now has to deal with an increased number of degrees of freedom; this can impact the overall duration of the optimization. For EPOS, when all the degrees of freedom are actively being optimized, it reaches a value of 1199. 

\subsection{\textit{EPOS optimization routine}}

\subsubsection{Initialization Parameters}

As mentioned before, EPOS aims to confine electrons and positrons in a high magnetic field. Additionally, it is important to explore the optimization landscape as thoroughly as possible. One choice that can affect this is the initial condition of the plasma; therefore, all the optimization runs start with a simple plasma shape, also known as a ``cold start". The LCFS shape is initialized with the coefficients $[a_{0,0}, a_{1,0}, a_{0,1}, a_{1,1}, d_{1,0}, d_{0,1}, d_{1,1}]$ being non zero and to have a minor radius of $\approx$ 4 cm. 

The coils are initialized before the optimization as circular coils with a major radius 10 \% larger than the plasma’s, and with a minor radius at 70 \% of the major radius of the plasma. However, additional care is required for the weave-lane coils which start with a major radius 20 \% bigger than the plasma's and a minor radius with the exact size of the major radius. 

Moreover, the currents flowing in the coils are calculated to achieve an on-axis field of 2.0 T for a fixed number of coils. The number of coils for EPOS is determined through a study on the degree of quasisymmetric error versus the number of coils for an EPOS-like configuration (see Appendix \ref{sec:num_coils}). The conclusion is that to achieve a satisfactory quasisymmetric configuration that is not affected by coil ripple, 10 coils plus one weave-lane are required. This number, however, is also a strict upper boundary because at low enough major radius, the absolute coil-to-coil distance becomes too small, increasing the engineering challenge.

Additionally, the plasma primary radius degree of freedom is kept fixed during all steps of optimization to ensure that it does not lead to a direct increase in the plasma volume nor bring the coils too close together.

\subsubsection{Optimization scheme}

\begin{figure} 
\centering
\begin{subcolumns}
    {\includegraphics[width=0.7\subcolumnwidth]{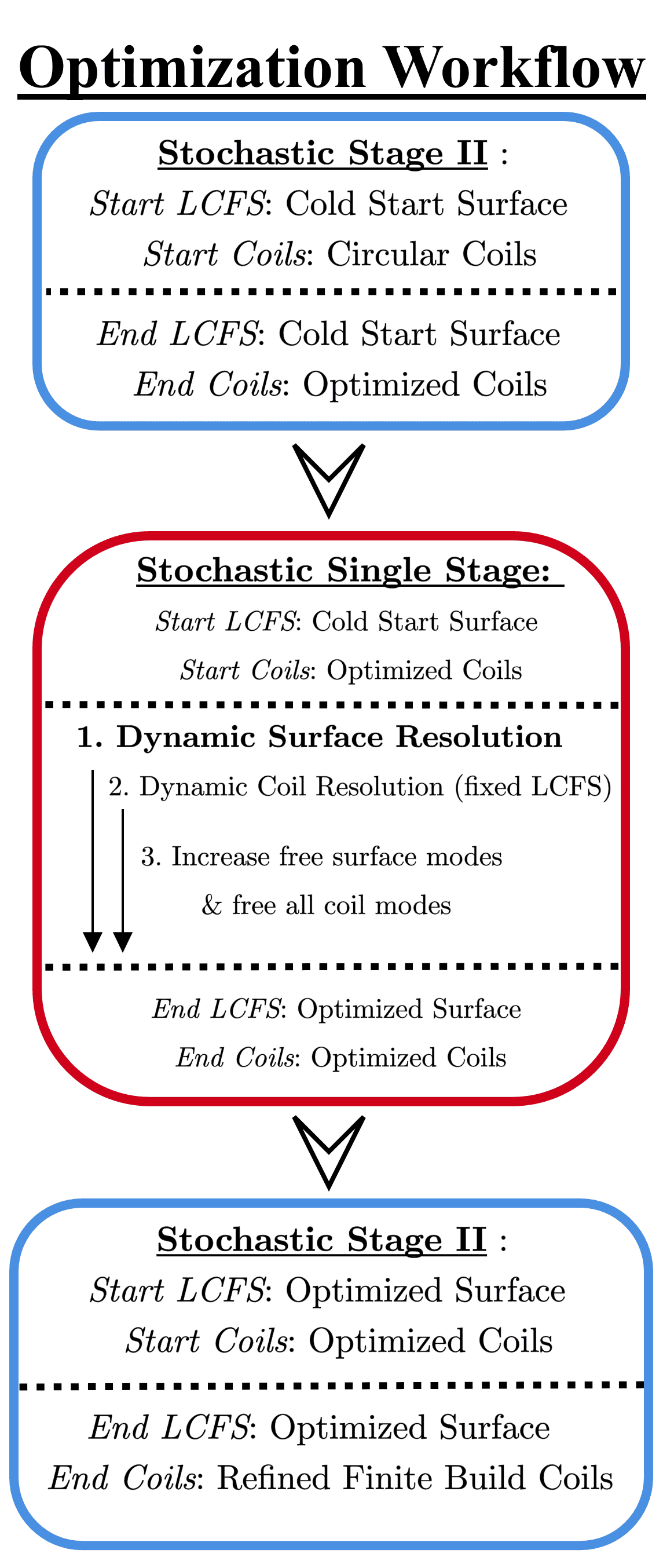}}
    \nextsubcolumn    {\includegraphics[width=\subcolumnwidth]{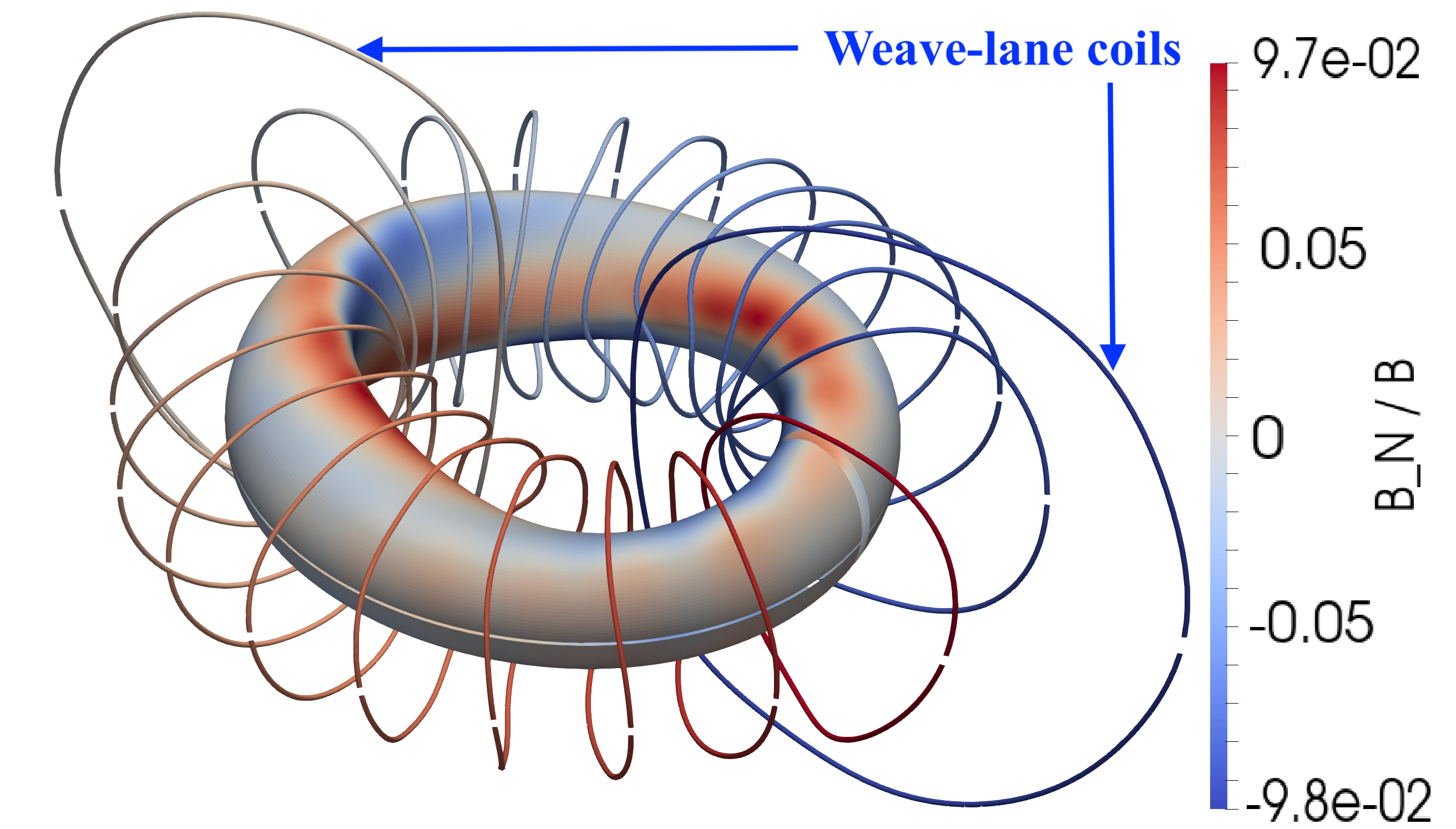}}
    \nextsubfigure
    {\includegraphics[width=\subcolumnwidth]{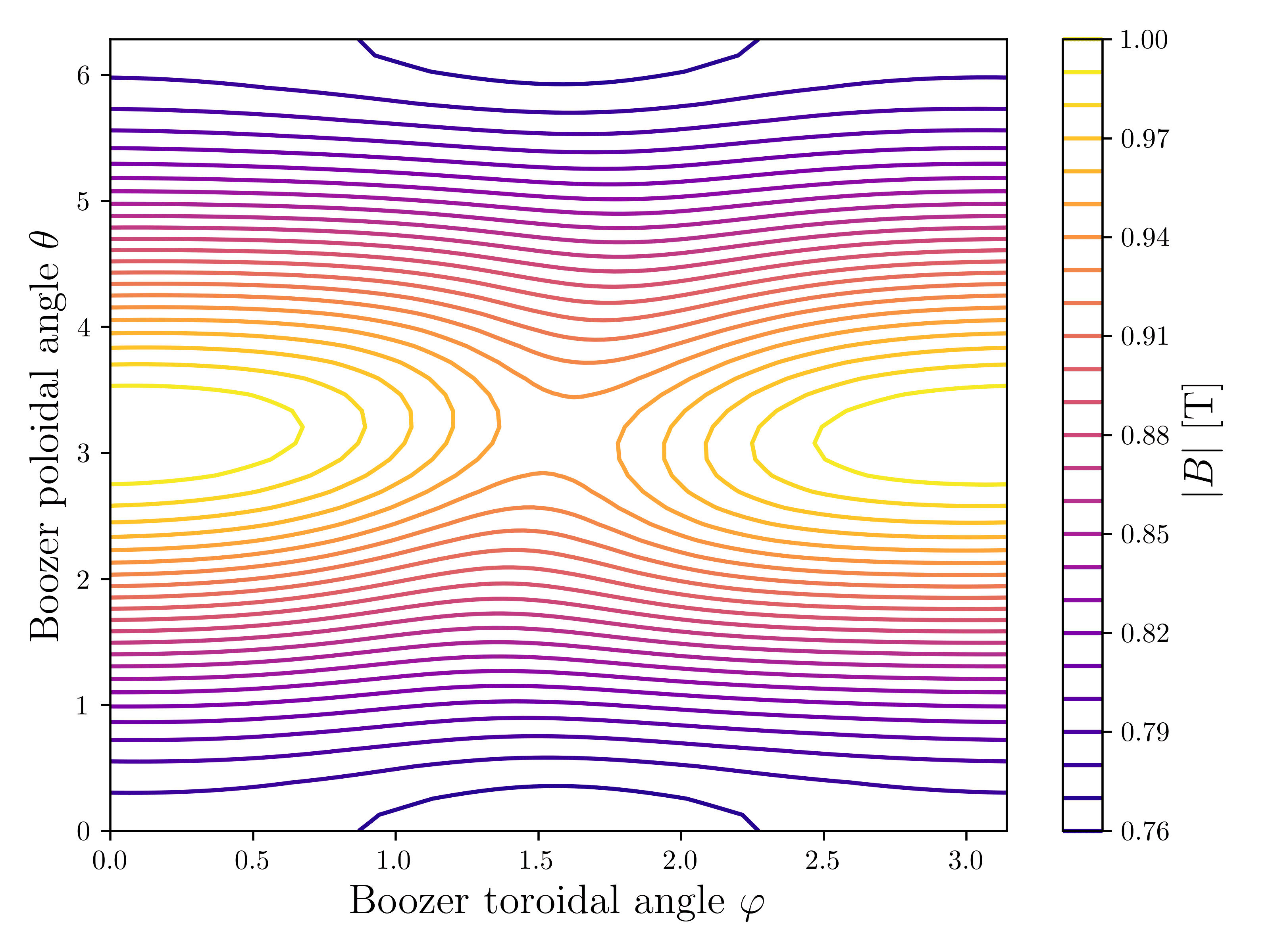}}
\end{subcolumns}
\caption{\textbf{Left:} Diagram displaying the optimization scheme for EPOS consisting of three main optimization stages: first, a stage II optimization is performed to find relatively accurate coils; followed by the stochastic single stage method focused on improving the coil shapes and optimizing the surface to balance physics and engineering constraints. Finally, stage II optimization is required to refine the HTS strain and coil concavity. \textbf{Top Right:} Rendering of the EPOS configuration with LCFS and coils after the single-stage step, the heatmap corresponds to the field error $\langle B\cdot n\rangle/\langle B \rangle$, note that its maximum value is purposely still relatively high as to not end up early on the design process in a sharp local minimum. \textbf{Bottom Right:} Boozer plot showing the quality of quasimmetry of the fixed-boundary ``cold start surface" used for the EPOS optimization.}
\label{fig:start_condition_and_stray_field}
\end{figure}

The goal of the optimization is to produce an end configuration that exhibits good physical properties, such as a flat iota profile and good quasisymmetry, and with adequate engineering solutions, including feasible and robust coils that accurately reproduce the magnetic field. In diagram \ref{fig:start_condition_and_stray_field}, the workflow to achieve this is presented with all the starting conditions mentioned in the previous section factored in.


The procedure starts by employing a stochastic stage II optimization similar to the work by \citet{Wechsung_2} to get initial coils that match the "cold-start" surface. This set of coils and surface can be visualized in Figure \ref{fig:start_condition_and_stray_field}. 
The objective function used for this preliminary optimization is the following: 

\begin{equation}
    f_{1} = w_{SF}\langle f_{SF} \rangle + w_L^{WL}f_L^{WL} + w_Lf_{L} + w_{bin}^{WL}f_{bin}^{WL}
    + w_{bin}f_{bin} + w_{tor}^{WL}f_{tor}^{WL} + w_{tor}f_{tor}.
\end{equation}


Then, using this configuration as a starting point, the stochastic single-stage is executed with the same plasma optimization technique used in stage I, sometimes referred to as dynamic surface resolution \citep{LandremanAndPaul}. Here, all the modes $[M_{pol}, N_{tor}]$ from Equation \ref{eq:surface} are fixed to start with, and then progressively one poloidal and one toroidal mode are unlocked after each iteration. Note that within one mode iteration, one complete optimization is performed. For example, in the first iteration, only the modes $m\leq 1$ and $|n| \leq 1$ are free. In this step, an analogous procedure is performed with the coil modes, which is referred to here as coil dynamic resolution. Once this phase is concluded, surface modes such that $m\leq 2$ and $|n| \leq 2$ are freed while keeping all the coil degrees of freedom also unfixed, and so on until all the coil degrees of freedom and all the surface degrees of freedom are unfixed. This method is adopted because it allows for a smooth transition of the coil shape.
In contrast, the surface shape cannot change considerably, which prevents overloading the optimizer with steep gradients in all directions. Freeing all the coil degrees of freedom for higher surface modes, which are unlocked, also enables sufficient flexibility in the coil shapes to match the increasing complexity of the surface as the dynamic resolution proceeds. In all the configurations presented in this paper, the full quadratic flux, as defined in Equation \ref{eq:squared_flux}, is targeted to be below 1e-4 after the single-stage step and to serve as a good starting point for the next stage. The plasma equilibria did not undergo a subsequent stage I after the single stage step, as they already fulfilled the quasisymmetry and iota profile requirements for EPOS. In order to reach a high enough resolution in the surface description to account for coil ripple, the LCFS is optimized up until $M_{pol} = N_{tor} = 12$, as defined in Equation \ref{eq:surface}, having in total 625 degrees of freedom as mentioned previously. The coil filaments are generated with $M_{coil} = 7$ from Equation \ref{eq:coils_fourier}, and $N_{\alpha} = 3$ in Expression \ref{eq:angle_fourier}, which corresponds to 574 degrees of freedom or a total of 1199, including surface and coils. The objective function for the single-stage step is presented as follows:

\begin{equation}
    \begin{split}
        f_{2} &= w_{QS}f_{QS} + w_{\iota}f_{\iota} +w_{A}f_{A} + w_{\Delta}f_{\Delta} + w_{S}f_{S} \\
        & + w_{SF}\langle f_{SF} \rangle + w_L^{WL}f_L^{WL}  + w_Lf_{L} \\
        &+ w_{bin}^{WL}f_{bin}^{WL} + w_{bin}f_{bin} \\
        &+ w_{tor}^{WL}f_{tor}^{WL} + w_{tor}f_{tor}.
    \end{split}
\end{equation}
During the optimization, the weight $w_{QA}$ is progressively increased as adding enough degrees of freedom to resolve coil ripple ($N_{tor} \geq 10$) proves to deteriorate quasisymmetry when not enough weight is added. The target width in the optimization is set to 4 cm for all the optimizations, as the volume should be minimized, and the target aspect ratio is set to 4.2. Moreover, the iota target is set to 0.2025, as values higher than 0.2 may indicate a high HTS strain and coil complexity. It should be noted that these three functions have low weights compared to the other terms, as they are meant to vary and remain relatively unconstrained during the optimization while still applying a constraint. The final step serves as a refinement step to ensure meeting the HTS strain engineering constraints, convex coils, finite build, and size of the weave-lane coils. This step is also stochastic, as all the previous ones, to ensure consistency in the methodology. The objective function used is:
\begin{equation}
    \begin{split}
        f_{3} &= w_{SF}\langle f_{SF} \rangle + w_L^{WL}f_L^{WL}  + w_Lf_{L} \\
        &+ w_{bin}^{WL}f_{bin}^{WL} + w_{bin}f_{bin} \\
        &+ w_{tor}^{WL}f_{tor}^{WL} + w_{tor}f_{tor} \\
        &+ w_{msc}^{WL}f_{msc}^{WL} + w_{msc}f_{msc}.
    \end{split}
\end{equation}
Here, the coils are modified to have finite dimensions in the optimization. This is achieved by simulating a winding pack with dimensions of 7 mm × 15 mm, consisting of 3 × 3 current-carrying filaments. Not only does it enable a higher fidelity calculation of the actual magnetic field, but it also provides a visual assessment of the tape's complexity, including its twists and bends. A considerable amount of fine-tuning of the weights is required in this final step. As guidelines the optimization is set to simultaneously meet a total squared flux smaller than $4\times10^{-6}$ $\text{m}^{2}$ using the local normalization method, an HTS strain in all the coils lower than 0.2 \%, negative curvature (for concavity reasons) no higher than -2 m$^{-1}$, and weave-lane coils no longer than 1.5 m and still displaying stray fields allowing for $\textbf{E}\times\textbf{B}$ injection. The value for the squared flux should be considered as an \textit{ad hoc} condition from previously made robustness verifications. These results showed that configurations with a squared flux higher than $6\times10^{-6}$ $\text{m}^{2}$ and coils perturbed with an amplitude higher than 0.5 mm do not yield sufficiently precise quadratic minimizing (QFM) surfaces with values as high as $6\times10^{-6}$ for the QFM residual.

\subsubsection{Definition of buildability}

The landscape of buildable stellarators meeting all the physics and engineering requirements mentioned previously is not trivial to find. In most optimization problems that are not inherently convex, there can be an extensive number of local minima. To explore this parameter space, two configuration scans are performed: one for the major radius and one for the ratio between the current of the weave-lanes and the one from the standard coils. Regarding the major radius, four values are explored: $[16,17,18,19]$ cm. Lower values are not considered, as they bring the coils too close together to preserve a low squared flux, and higher values are ruled out since they naturally require more positrons. For the weave-lane current ratio, two ratios are considered: $[3,4]$, a ratio of 2 is not considered, as the optimizer consistently reduces the size of the weave-lane coils in order to approximate as much as possible a stellarator symmetric device, hence preventing the presence of stray fields for weave-lane coils with small dimensions. Therefore, in this paper, eight different configurations are presented, all of which can be designed and built. All the stochastic optimizations are performed with an absolute perturbation amplitude $\sigma$ of 1 mm, so as to overshoot the precision requirements of 0.5 mm in the trenches that are expected at the current engineering state. This is to account for additional tolerances that have to be taken into account due to assembly inaccuracies, thermal loads, and electromagnetic forces. These perturbations are done with a characteristic length $L_s = 0.2$ m, given that most of the coil deviations will originate from the slack in the winding pack trenches, and higher frequency perturbations should not arise due to the tape's stiffness in the binormal and normal directions.

Moreover, the WL stray fields should correspond to a group of field lines extending from the plane of the weave-lanes, crossing a boundary set by a cylinder of 60 cm in radius and about 80 cm high corresponding to the vacuum chamber walls, and reaching a region of 5 mT outside the chamber, taken as the standard field strength of guiding tubes necessary to inject the positrons, this is shown in Figure \ref{fig:start_condition_and_stray_field}. Another aspect considered in the optimization and subsequently checked for all candidates is the presence of positive normal curvature. Normal curvature in the rotated centroid frame is defined as :

\begin{equation}
    \kappa=\frac{\Tilde{\mathbf N} \cdot \Tilde{\mathbf T}^\prime}{\|\boldsymbol \Gamma'\|}.
\end{equation}
Where $\Gamma$ describes the curve in 3 dimensions, $\Tilde{\mathbf T}$ is the tangent vector to the curve, and $\Tilde{\mathbf N}$ is a modified normal curvature vector (see \cite{huslage2024strainoptimizationrebcohightemperature}). This quantity is positive in concave sections of the coils. The tape will be similarly wound around the trenches, just like magnetic tape is wound on a spool. This means that any positive curvature sections would make the winding procedure considerably more complicated, as no tension can be sustained throughout the process. 

The simulations are all performed on the VIPER cluster of the Max-Planck Computing and Data Facility, which comprises a total of 768 nodes featuring AMD EPYC Genoa 9554 CPUs with 128 cores and at least 512 GB of RAM per node. Stochastic optimization depends on an MPI implementation of the codes running on five nodes with 128 cores each, where one core executes the code and computes the results for one sample. Together, the optimization uses 1280 samples, and 350 GB of RAM is reserved for these purposes.

\section{Numerical Results}
\label{sec:results}
\subsection{EPOS Candidates}

The investigation of potential designs for the EPOS is explained here, with eight configurations of both coils and equilibrium. They are labeled as C(3,4)\_R(16,17,18,19), where C refers to the standard coil to weave-lane coil current ratio and R to the major radius of the configuration. The analysis presented in this paper is conducted to the extent necessary to allow for a first design freeze of the configuration. A subsequent engineering design phase is necessary to assess the feasibility of the stellarator fully. This will include, for example, inspections of the coil stresses, magnetic permeability of materials, and their error fields with finite-element simulations. If any modifications are needed, other configurations may be considered or modified.
The metrics that are shown here are calculated from the reconstructed VMEC equilibria generated by the coils. This can be achieved by optimizing a QFM surface for the last closed flux surface to fit the one actually produced by the coils. Then, by updating the boundary and rerunning VMEC, the information for the vacuum equilibrium can be obtained. The optimization for all cases is performed using 16 poloidal modes and 22 toroidal modes, with a tolerance error of $10^{-12}$, sufficient to resolve the coil ripple of the magnetic field lines, which can significantly damage the quasisymmetric error (QS error). The coils used in the reconstruction are generated with nine filaments per coil to simulate their finite dimensions as described previously.

\begin{figure} 
    \centering
    \includegraphics[width=0.47\linewidth]{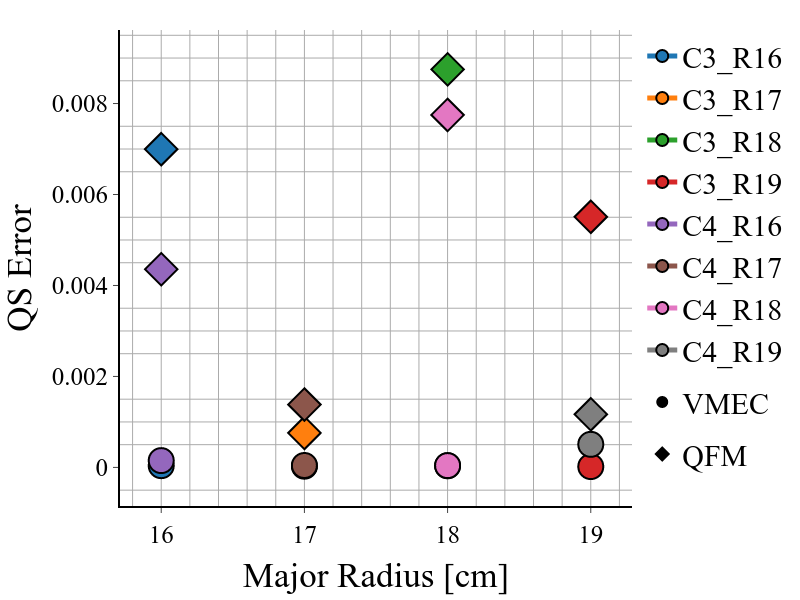}
    \hspace{0.5cm}
    \includegraphics[width=0.47\linewidth]{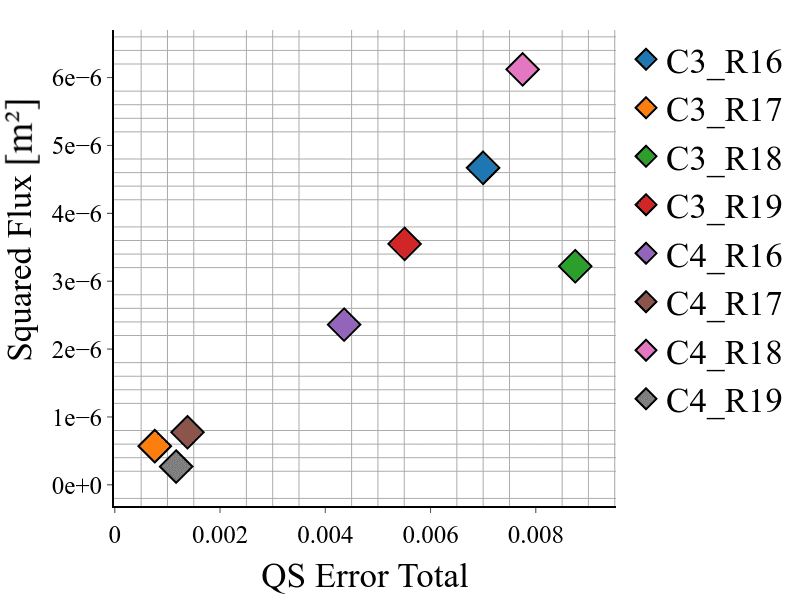}

    \includegraphics[width=0.47\linewidth]{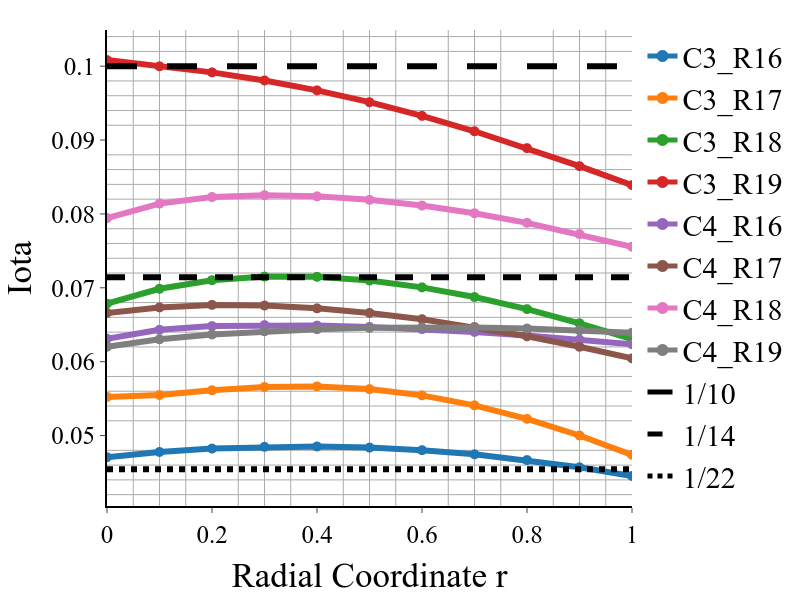}
    \hspace{0.5cm}
    \includegraphics[width=0.47\linewidth]{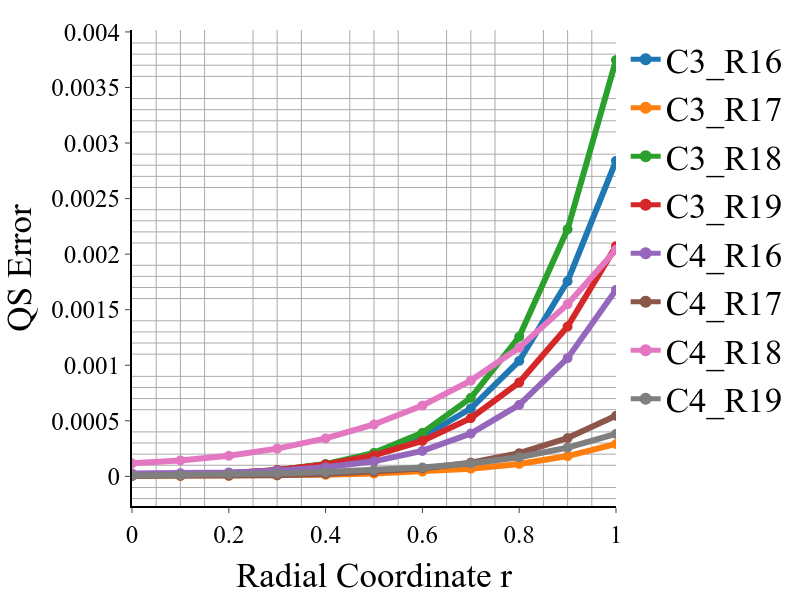}
    \caption{\textbf{Top Left:} Quasisymmetric error of the EPOS configurations plotted against the major radius for both the VMEC equilibrium and the QFM-generated one. \textbf{Top Right:} Squared Flux vs Quasisymmetric error for all configurations. \textbf{Bottom Left:} Iota profiles for all the EPOS candidates and the three most important rationals bounding the rotation transform values. \textbf{Bottom Right:} Quasisymmetric error radial profile for the EPOS configurations}
    \label{fig:qs_data}
\end{figure}

In Figure \ref{fig:qs_data}, both the total quasisymmetric error and the total squared flux as a function of the quasisymmetric error are shown. The prior is the sum over 11 flux surfaces, and the total QS error is plotted for both the VMEC and the QFM values. Here, VMEC circle data points correspond to the target QS error value from the VMEC equilibrium object that the optimizer obtained in the single-stage step, which is not the one generated by the coils. In contrast, the diamond QFM data points represent the QS value from the reconstructed equilibria from the final coils. The first observation is that all the QFM values are higher than the VMEC ones. This is expected, as they primarily come from coil ripple, given that there is a finite number of coils with finite dimensions. A considerable variation in the QS quality of the QFM configurations is also visible, as the configurations are optimized to account for engineering constraints, which makes the choice of weights unpredictable and results in very different configurations, rather than just scaled-up or down versions of each other. However, for 16, 18, and 19, all coil sets with a current ratio of 4 performed better than their counterparts with a current ratio of 3. C3\_R17 can be considered an exception, as it is optimized with the purpose of disregarding concavity in the coils, ultimately allowing it to reach regions of high accuracy in the optimization space more easily. 

Moreover, as shown on the top right plot of Figure \ref{fig:qs_data}, the better the total squared flux, the better the quasisymmetry of the considered configuration. Note that the correlation is not one-to-one, as local variations of the field generated by the coils on the target magnetic surface are only on average detrimental to the quasisymmetry, and a certain percentage of these can amount to a local improvement of quasisymmetry. However, the most accurate configurations are those with the lowest squared flux, which can result in an almost one order of magnitude better QS metric.
Another aspect to consider when designing a stellarator is the iota profile. As mentioned previously, for EPOS, a flat iota profile value between 0 and 0.1 is targeted. As shown in the bottom left of Figure \ref{fig:qs_data}, all the configurations besides C3\_R19 possess an iota that varies at most 10 \% radially. The highest ratios that bound the iota profiles, except for C3\_R19, are 1/14 and 1/22, meaning that no major island chains are expected to appear in the magnetic field. 

\begin{figure} 
    \centering
    \includegraphics[width=0.32\linewidth]{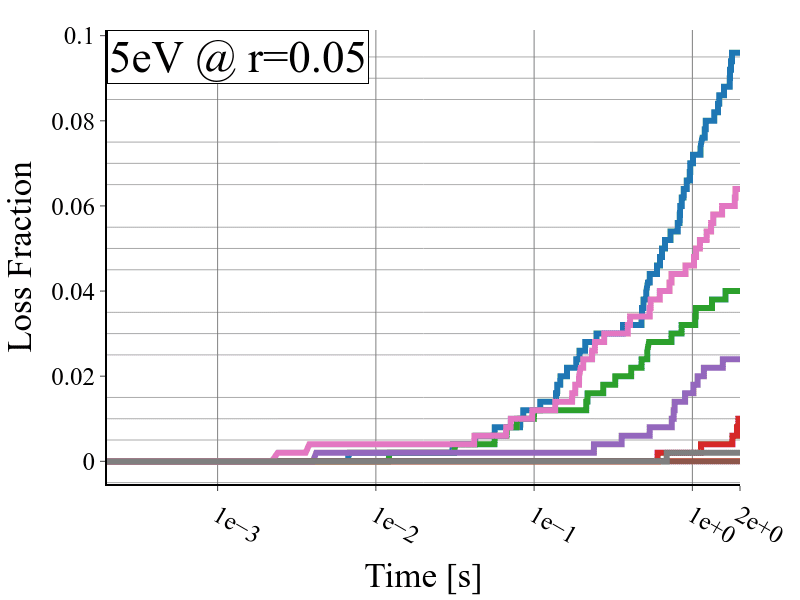}
    \includegraphics[width=0.32\linewidth]{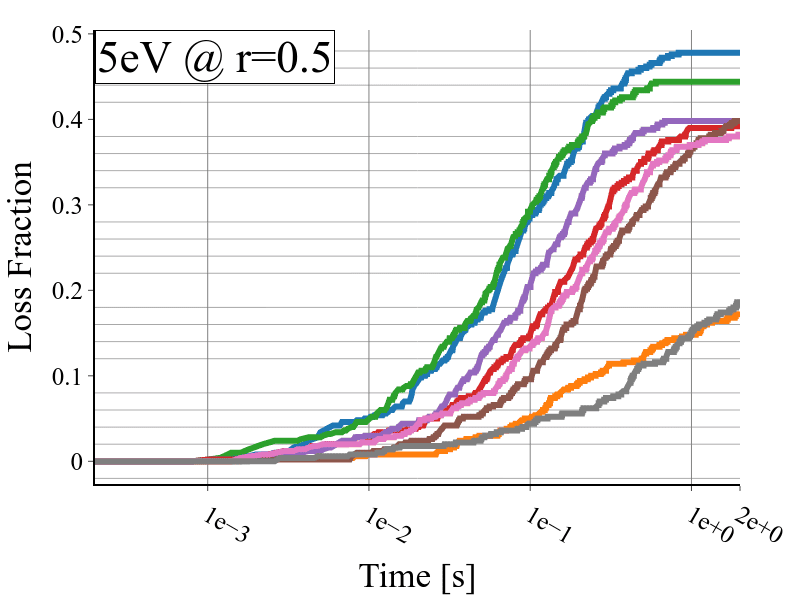}
    \includegraphics[width=0.32\linewidth]{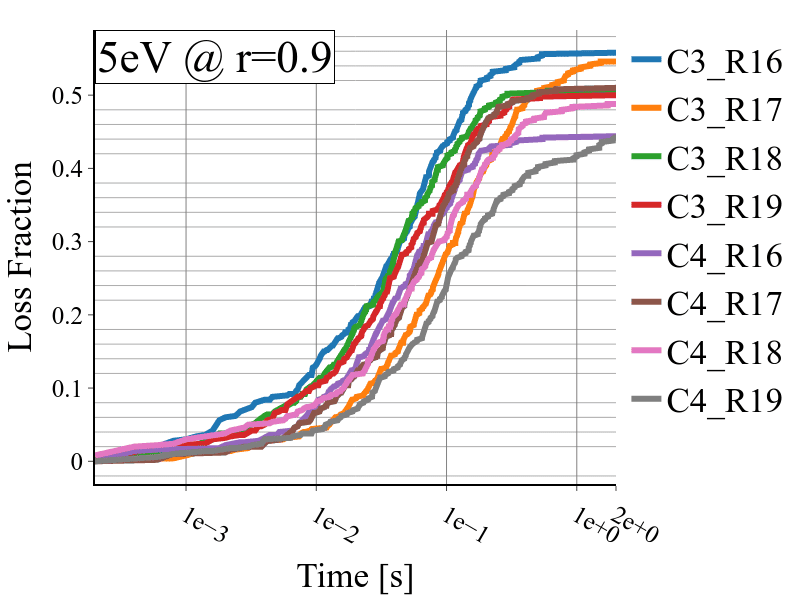}
    \caption{Loss fraction of positrons at 5 eV versus time of the EPOS configurations with starting conditions at s=0.05 (\textbf{Left}), s=0.5 (\textbf{Middle}), and s=0.9 (\textbf{Right}). Here we show that in the worst case scenario with no cyclotron cooling, at 5 eV, we are able to keep almost 100 \% of the particles after 2 seconds when starting the particles at the magnetic axis. This ratio reduces to about 60 \% when starting at the outer boundary.}
    \label{fig:confinement_data}
\end{figure}

The fact that the iota of all the configurations remained relatively low compared to other known devices, such as W7-X, allows EPOS to find simpler coils with lower binormal and torsional strain that can potentially damage the coils during winding, for example. 
Then, although the possible values for the major radius are between 16 and 19 cm and the target aspect ratio is 4.2, the optimizer kept all of the aspect ratios below 4.2; on top of that, the bean cross-section width is also penalized with a target value of 4 cm, making it more compact than anticipated. This is related to the fact that the weights of the aspect ratio and width are purposely kept low in comparison to other terms, allowing more freedom for the gradient descent to find suitable minima. 
The variation in QS error can be further observed in the profiles on the bottom right of Figure \ref{fig:qs_data} where three configurations stand out as particularly accurate: C3\_R17, C4\_R17, and C4\_R19 with values all lower than $10^{-3}$ throughout the volume, which is coherent with the results displayed on the top left of Figure \ref{fig:qs_data}. Moreover, as expected, all configurations perform substantially better as they approach the magnetic axis. This suggests that, regardless of the configuration, core injection should be considered whenever possible to maximize the confinement time of positrons and electrons.
This is supported by the data displayed in Figure \ref{fig:confinement_data}. The data is produced using the code SIMPLE \citep{Albert_Kasilov_Kernbichler_2020,SIMPLE_albert}, which uses symplectic integration in systems where the Hamiltonian is specified in non-canonical coordinates for the estimation of alpha particle losses of guiding-center orbits in a 3D geometry, but adapted for the loss of positrons. The simulations are performed with 500 particles with 5 eV and a random distribution of $v_{\perp}/v$ and at different radial coordinates: 0.05 (at the magnetic axis), 0.5, and 0.9 (near the LCFS). This is done to inform on the confinement performance of all the configurations depending on the injection scheme. 

The time duration of the particle simulation is 2 seconds, which is estimated to be long enough to radiate about 40 \% of the perpendicular energy through cyclotron cooling \citep{anderegg_cyclotron_cooling}. This ensures that most of the particles possess mainly parallel energy, i.e., are passing particles and therefore can be indefinitely confined assuming flux surfaces and no other transport effects. It is visible that starting the simulations at the core yields, in the worst case scenario, a loss of only 10 \% of particles after 2 seconds and, in the best case, 0 \% losses. This confinement is however degraded as the particles are injected closer to the edge with losses reaching 50 \% with two outliers (C3\_R17 and C4\_R19) at less than 20 \%. Finally, for all configurations, results that start with the particles close to the LCFS simulating an edge-injection type scenario reach between 55 \% and 40 \% losses. These results illustrate the quality of the confinement/quasisymmetry of the configurations; however, little can be concluded about the injection scheme for EPOS, as it will need to be fine-tuned experimentally primarily. Moreover, the number of confined particles of the built machine will be degraded to some extent due to injection inefficiencies and manufacturing and assembly errors, which will further impact the quality of quasisymmetry.

\begin{figure} 
    \centering
    \includegraphics[width=0.47\linewidth]{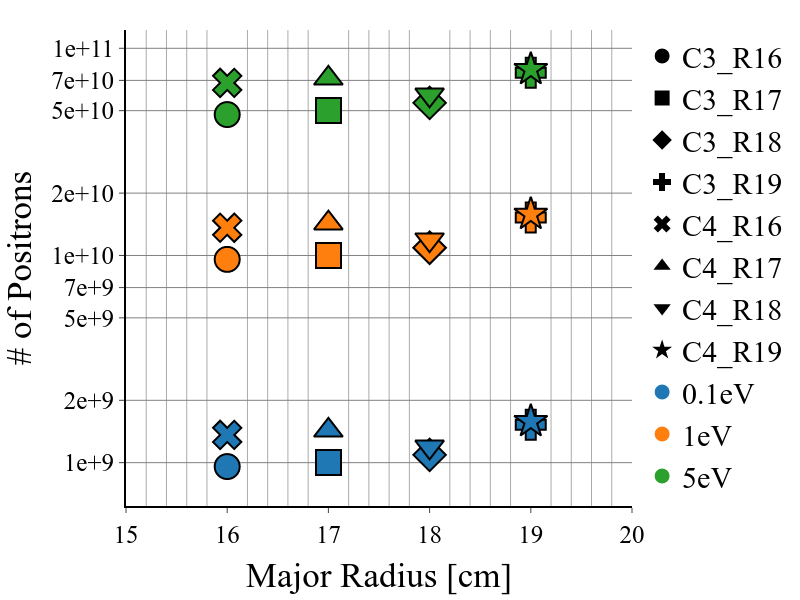}
    \hspace{0.5cm}
    \includegraphics[width=0.47\linewidth]{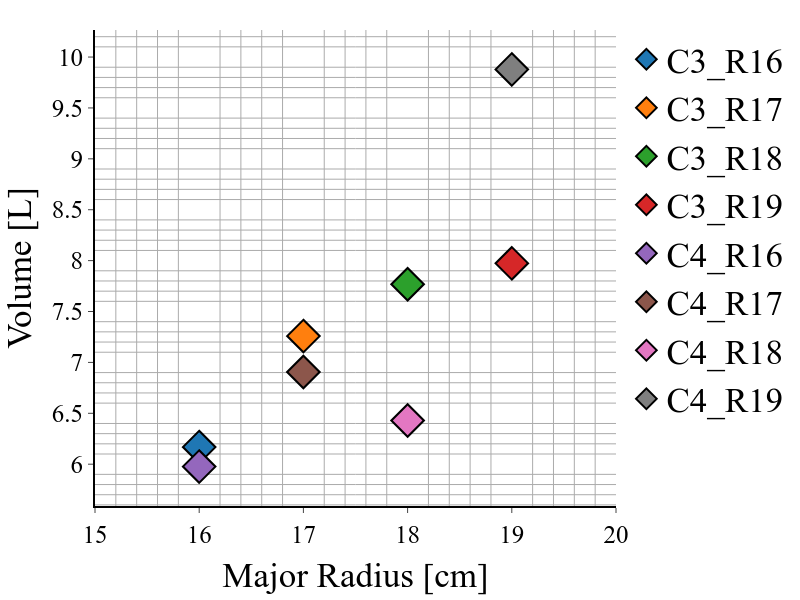}
    \caption{\textbf{Left:} Minimal number of positrons required to read $a/\lambda_{D} \geq 10$ for all the configurations as a function of the major radius and plotted for different temperatures (0.1, 1, and 5eV).\textbf{Right:} Volume taken by the vacuum field of all the configurations as a function of the major radius.}
    \label{fig:positrons_number}
\end{figure}

Moreover, the main idea of the scan is driven by the scaling of $a/\lambda_{D}$ as a function of the major radius, as shown in Equation \ref{eq:debye_length}, where a smaller major radius means a higher ratio and therefore it reduces the amount of required positrons. In Figure \ref{fig:positrons_number} is shown the minimal amount of necessary positrons to meet the requirement of $a/\lambda_{D} \geq 10$ for all the EPOS configurations as a function of the major radius and for three different positron temperatures, namely 0.1, 1, and 5 eV. It is visible that the temperature is the parameter that primarily impacts the number, and that the major radius of the configurations does not impact as much. Another quantity that can directly impact the $a/\lambda_{D}$ ratio is the volume (shown to the right in Figure \ref{fig:positrons_number}). As expected the volume increases with the major radius of the configurations, meaning that the Debye Length increases. Since there is such a high variability in parameters such as the minor radius and hence in the aspect ratio, as supported by Figure \ref{fig:positrons_number}, this can explain the minor enhancement of the necessary particle number and simultaneously means that the number of required positrons of a configuration is no longer a primary parameter in the choice of the stellarator. 

\begin{figure} 

\begin{subcolumns}
{\includegraphics[scale=0.26]{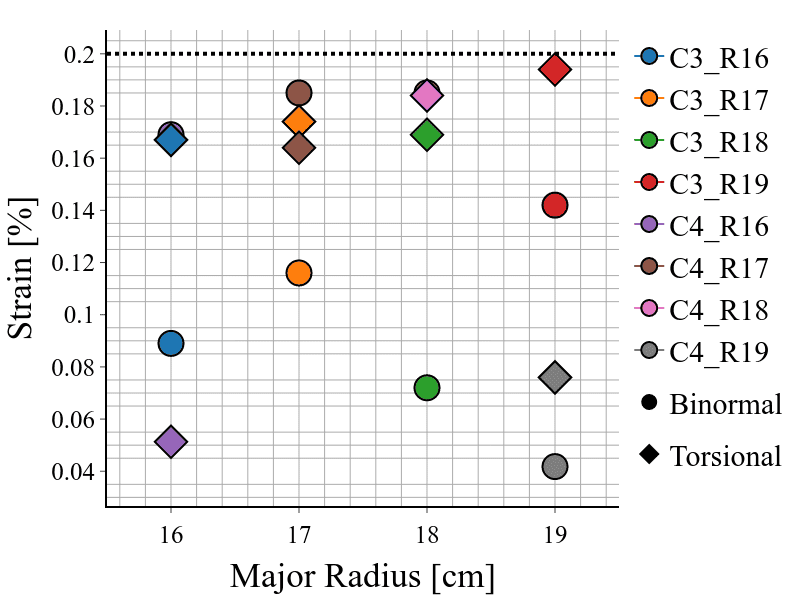}}
\nextsubcolumn
{\includegraphics[width=\subcolumnwidth]{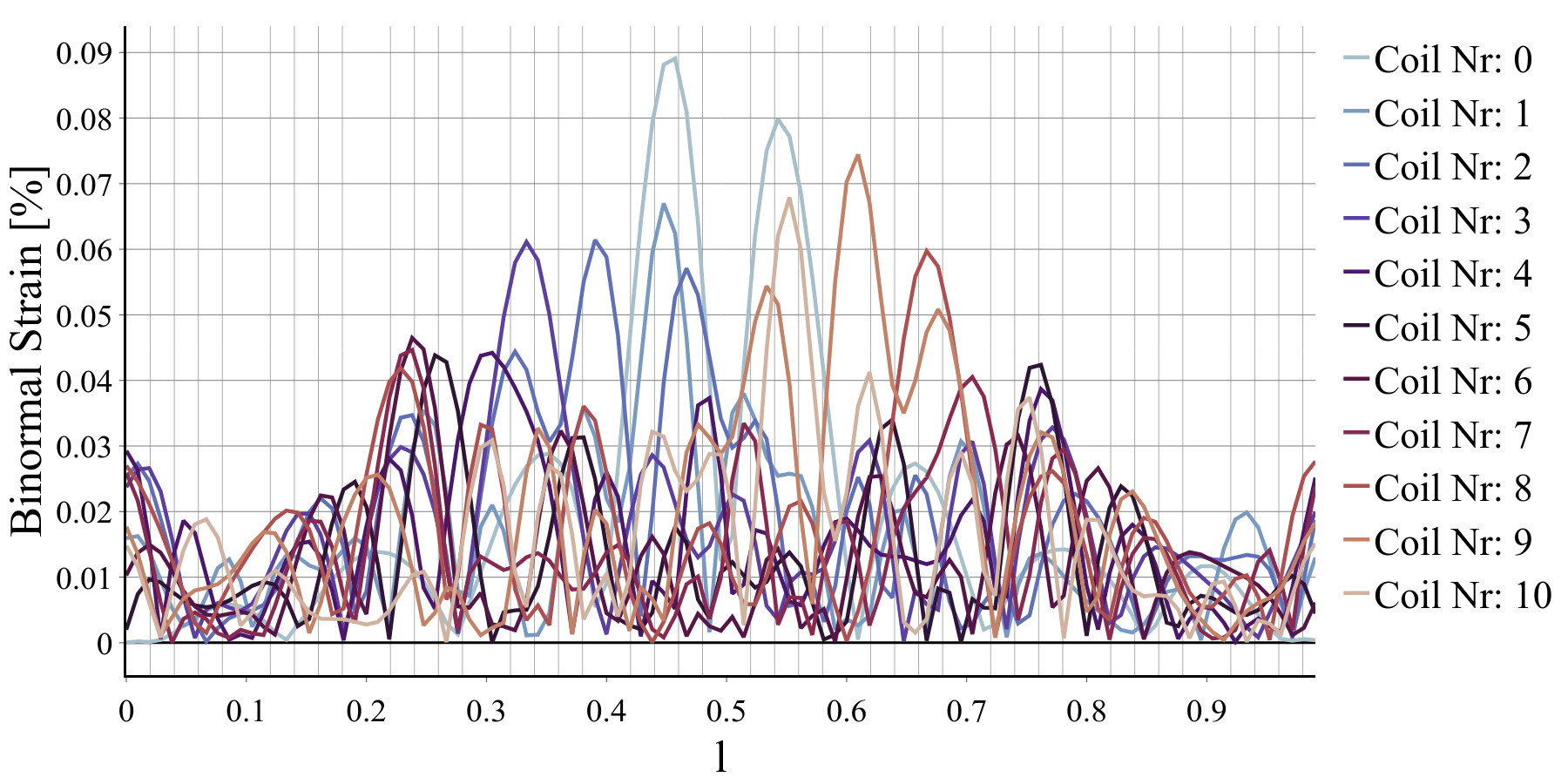}}
\nextsubfigure
{\includegraphics[width=\subcolumnwidth]{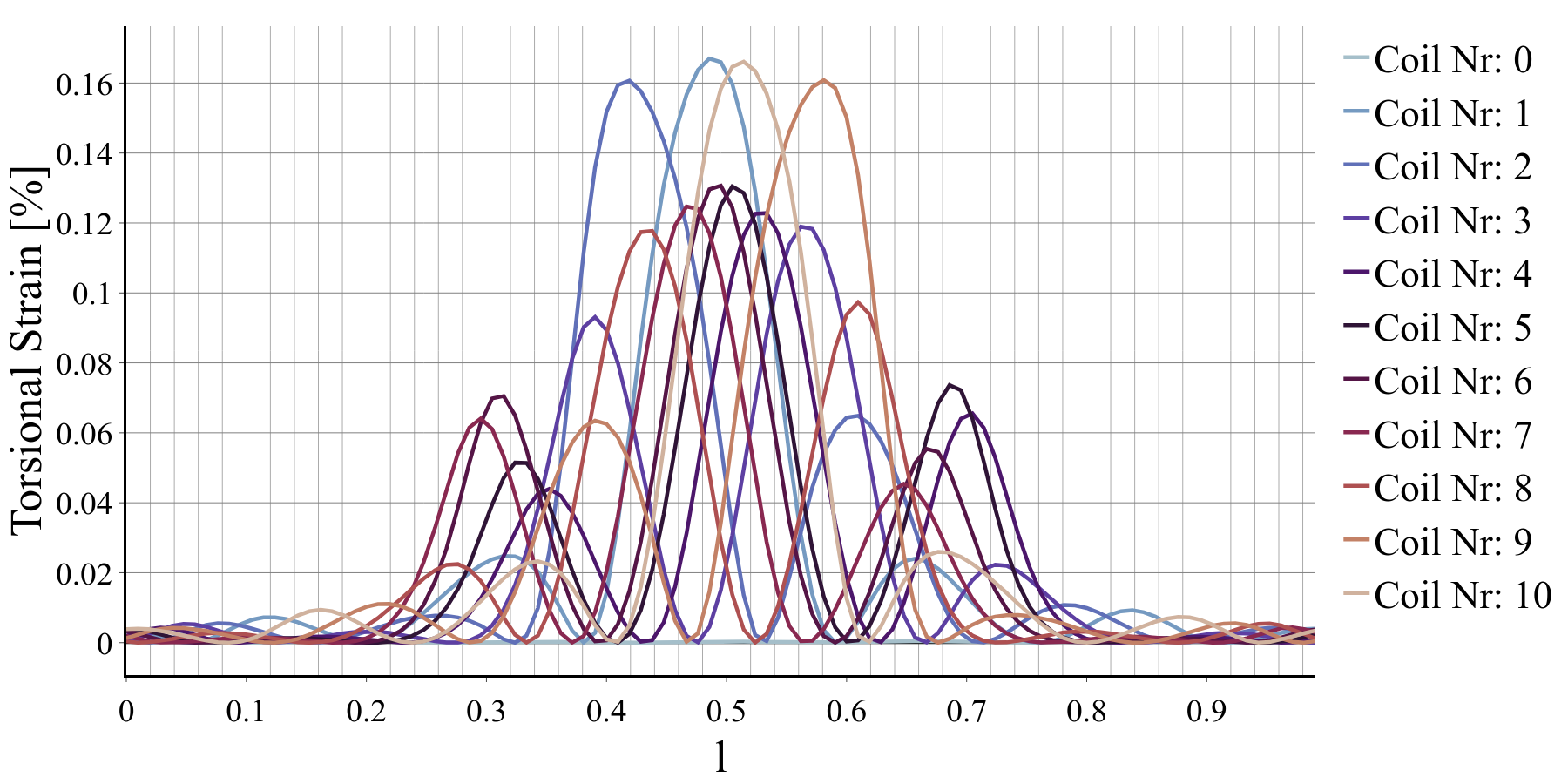}}
\end{subcolumns}

\caption{\textbf{Left:} Maximum HTS Torsional and Binormal strain of all the configurations presented. \textbf{Top Right:} HTS Binormal strain of all the coils of the C3\_R16 configuration. \textbf{Bottom Right:} HTS Torsional strain of all the coils of the C3\_R16 configuration.}
\label{fig:HTS_strain}

\end{figure}

\begin{figure} 
    \centering
    \includegraphics[width = 0.45\linewidth]{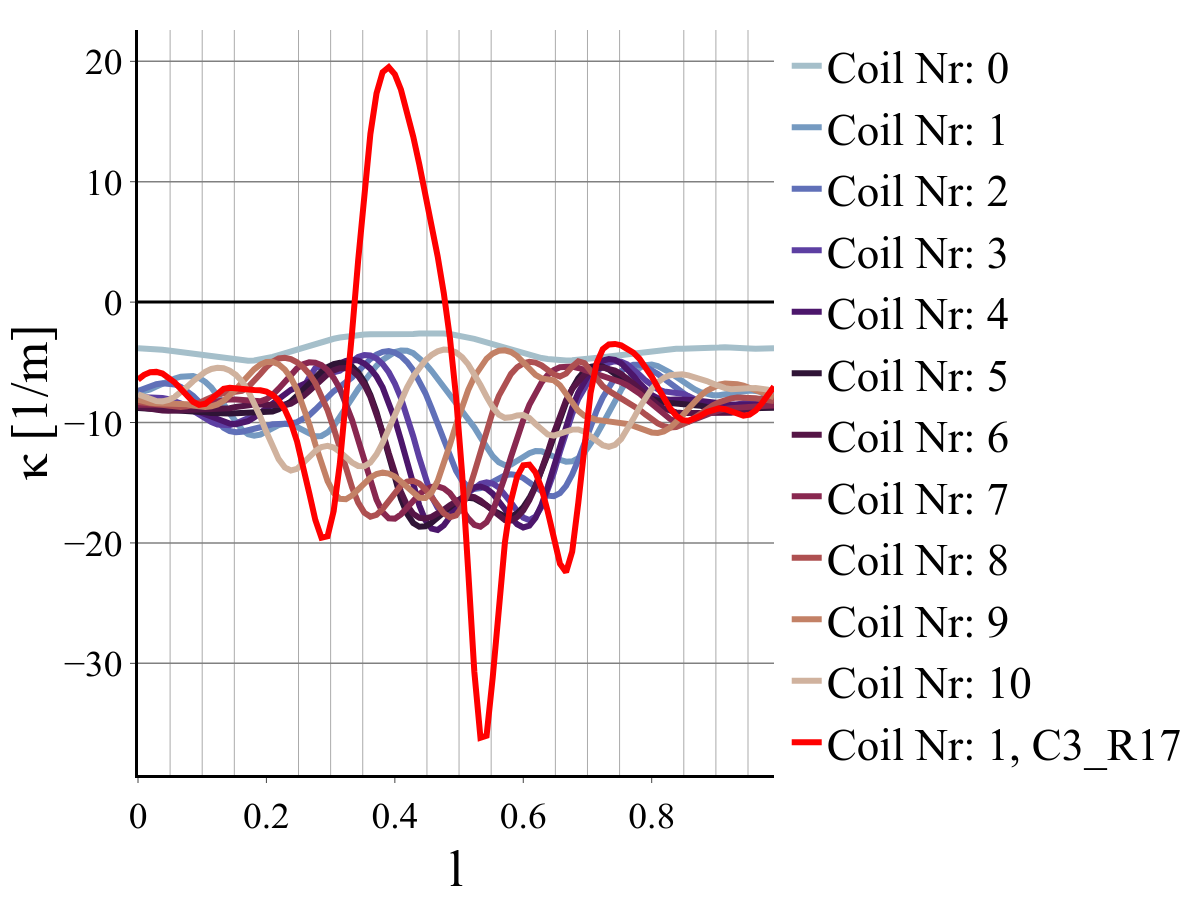}
    \hspace{0.5cm}
    \includegraphics[width = 0.35\linewidth]{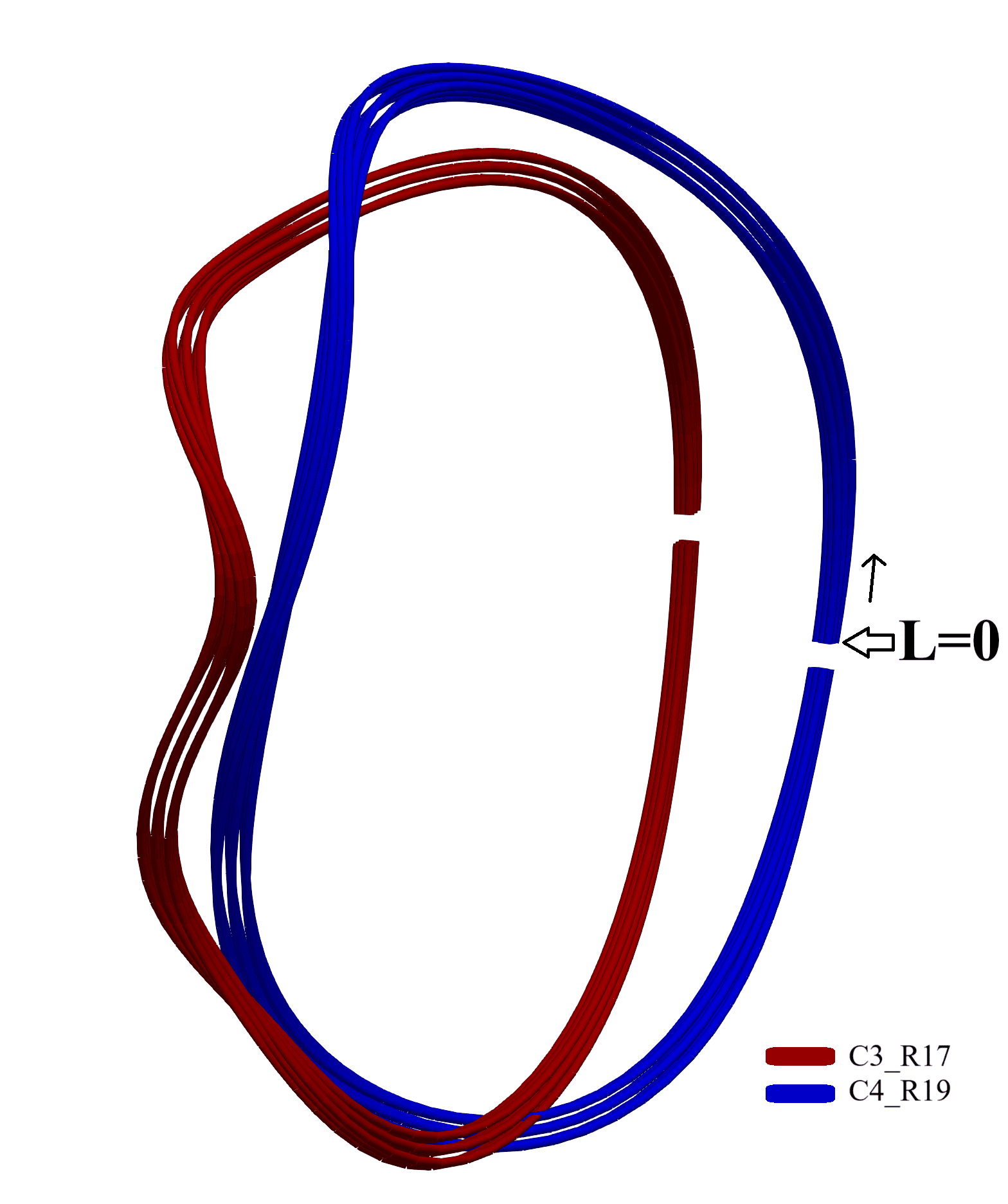}
    \caption{\textbf{Left:} Normal curvature profiles for all the coils of configuration C4\_R19 and for coil 1 of configuration C3\_R17 as an example of highly concave sections. \textbf{Right:} Rendering of coil 1 of C3\_R17 showing the concave section and of coil 1 of C4\_R19 demonstrating that simple convex coils are achievable as well.}
    \label{fig:concave_coils}
\end{figure}

Designing buildable coils has two main goals: simplification of the manufacturing process, by minimizing the development of winding machines and winding pack technology, for example, and ensuring that the designed coils can be replicated within reasonable accuracy. This is because extremely twisted and complex coils are inherently more difficult to wind. Therefore, it is additionally laborious to accurately shape them, as well as to verify that they comply with tolerance requirements.
In Figure \ref{fig:HTS_strain} are shown the maximum torsional and binormal strain values of all the configurations. All the values are below the 0.2 \% threshold, which is already a conservative value for the maximal accepted strain value, while simultaneously showing acceptable squared fluxes that all score below $6\times10^{-6}$ $\text{m}^{2}$ as shown in Figure \ref{fig:qs_data}. An example of the strain profiles of what is considered a safe coil to build is shown on the right in Figure \ref{fig:HTS_strain}. These profiles are characteristic of QA for EPOS, as the inboard side (at $l \approx 0.5$) requires the most shaping of the coils, which is evident in both torsional and binormal strain. The shaping can also be observed in the curvature $\kappa$ profiles and the physical coils of Figure \ref{fig:concave_coils}. The areas of high coil complexity are, therefore, in the EPOS case, very localized.

Moreover, EPOS will be built using non-insulated coils. As highlighted earlier, concave sections prevent the tension from being kept on the tape while winding. Tension is necessary as it establishes good thermal and electrical contacts between HTS layers in the winding packs. Thanks to the subsequent current and heat redistribution in case any section of the HTS becomes normal conducting, the coils become passively robust to quenches, which is why non-insulated (NI) coils are of such interest in magnet research \citep{NI_coils,NI_coils2}. Therefore, a requirement for EPOS is that all the coils are convex. In Figure \ref{fig:concave_coils}, the standard curvature profiles along each coil of configuration C4\_R19 and of coil 1 of the set C3\_R17 are shown. The curvature of all the coils of the former stellarator is negative, indicating a consistent convex structure of the coil, which is also visible in the 3D model of coil 1 of the respective stellarator. As an example of what the optimizer systematically finds as a stable minimum, the curvature of coil 1 of the C3\_R17 is also plotted, displaying very strong concavity around the inboard side of the coil, reaching a value of 20 $\text{m}^{-1}$. To understand its concavity, the coil is also shown on the right. A visual difference between the two coils is that the C3\_R17 possesses the characteristic 'S' shape on the inboard side that gives rise to the rotational transform, which, together with a minimization of the torsional and binormal strain, enhances the concavity of the coil. It is nevertheless possible to generate a rotational transform without twisting the coils to such complex levels and without breaking the HTS strain limits. This method will be explained in the following section.

\begin{figure} 
    \centering
    \includegraphics[scale = 0.4]{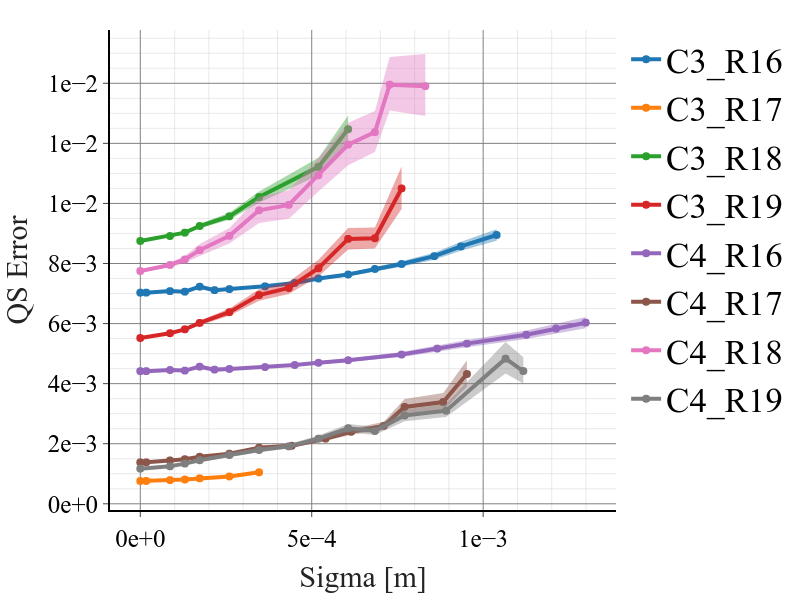}
    \caption{Robustness of the quasisymmetry versus coil pertubration amplitude for all the configurations. Here, the metric is the average deviation from quasisymmetry at a certain perturbation amplitude of the coils.}
    \label{fig:robustness_data}
\end{figure}

Another essential parameter when considering building an EPOS-sized device is the manufacturing tolerances. This is the main reason why stochastic optimization is used throughout the process. In this section, an a posteriori study, meaning after all the optimization steps, is performed to check the robustness of the many configurations already presented previously. The method is very similar to the generation of perturbed samples in the optimization; however, here the coils that are perturbed are of finite dimension with nine filaments. Each filament of a single coil is perturbed in the same way to simulate deviations of the whole winding pack. The perturbations are done using systematic deviations that are present in coils of the same type and statistical errors that are different in every coil. Then, a QFM surface is calculated for each sample, and a VMEC equilibrium is generated to extract data on the perturbed quasisymmetry from it. This is performed for all samples, and for increasing perturbation amplitude $\sigma$, a one-sigma data point is only added when all the VMEC samples converge. It should be noted that $\sigma$ represents the pre-exponential factor of the kernel in the Gaussian Process along one dimension; in order for it to correspond to a 3D perturbation, $\sqrt{3}$ should be factored in. Each sampling is performed with 128 samples. The results are shown in Figure \ref{fig:robustness_data}.

There are two visible categories: in the first one are configurations C3\_R18, C3\_R19, and C4\_R18, for which the departure from quasisymmetry increases relatively fast with increasing perturbation amplitude compared to the rest of the configurations, which appear to be robust to any deviations up to more than 1.3 mm for the C4\_R16 stellarator, without any significant deterioration of the QS quality. C3\_R17 appears as an outlier because although its quasisymmetric error is low as well as its relative increase, for deviations higher than 3.5 mm, it is impossible to get all the samples to converge, indicating that this configuration might find itself in a sharp local minima. This is possible since this is the only configuration where high curvature is allowed, thereby minimizing the squared flux as much as possible. According to the Figure \ref{fig:robustness_data}, this results in a configuration that is highly accurate but not robust. This, together with configuration C3\_R16, is one of the two configurations that show that the squared flux alone is not enough to quantify the robustness of a stellarator, as it is intuitive to think that a highly accurate stellarator can allow more damage to its coils as it compensates for the loss of accuracy. Moreover, the data displayed in Figure \ref{fig:robustness_data} indicates that there is an inherent robustness of a stellarator. It is unclear at this stage what the decisive factor behind it is. An overview of the critical parameters of all the candidates presented here is shown in Section \ref{sec:spider_plots}. 

\subsection{Refining the Optimization}
\begin{figure} 
    \centering
    \includegraphics[width=0.47\linewidth]{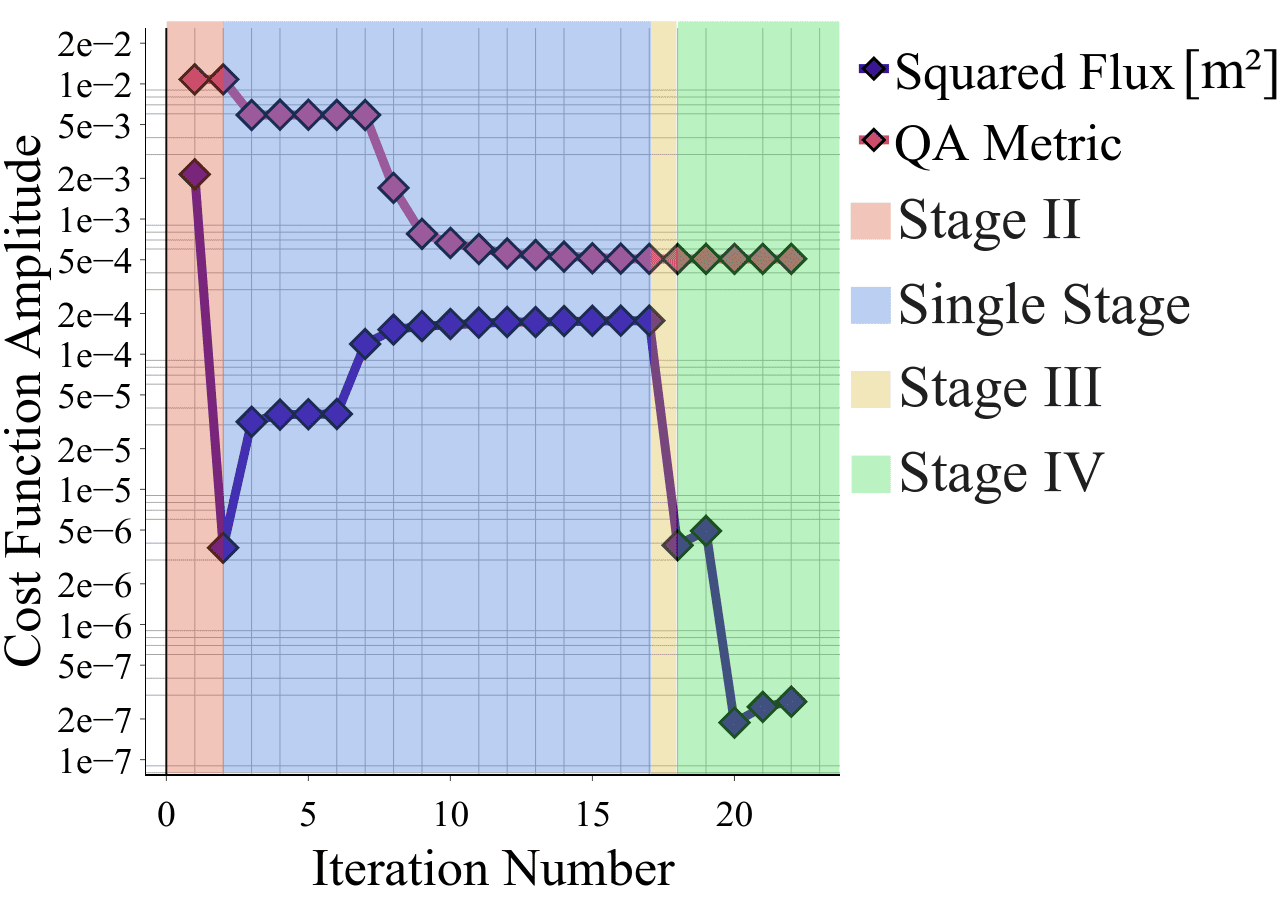}
    \includegraphics[width=0.47\linewidth]{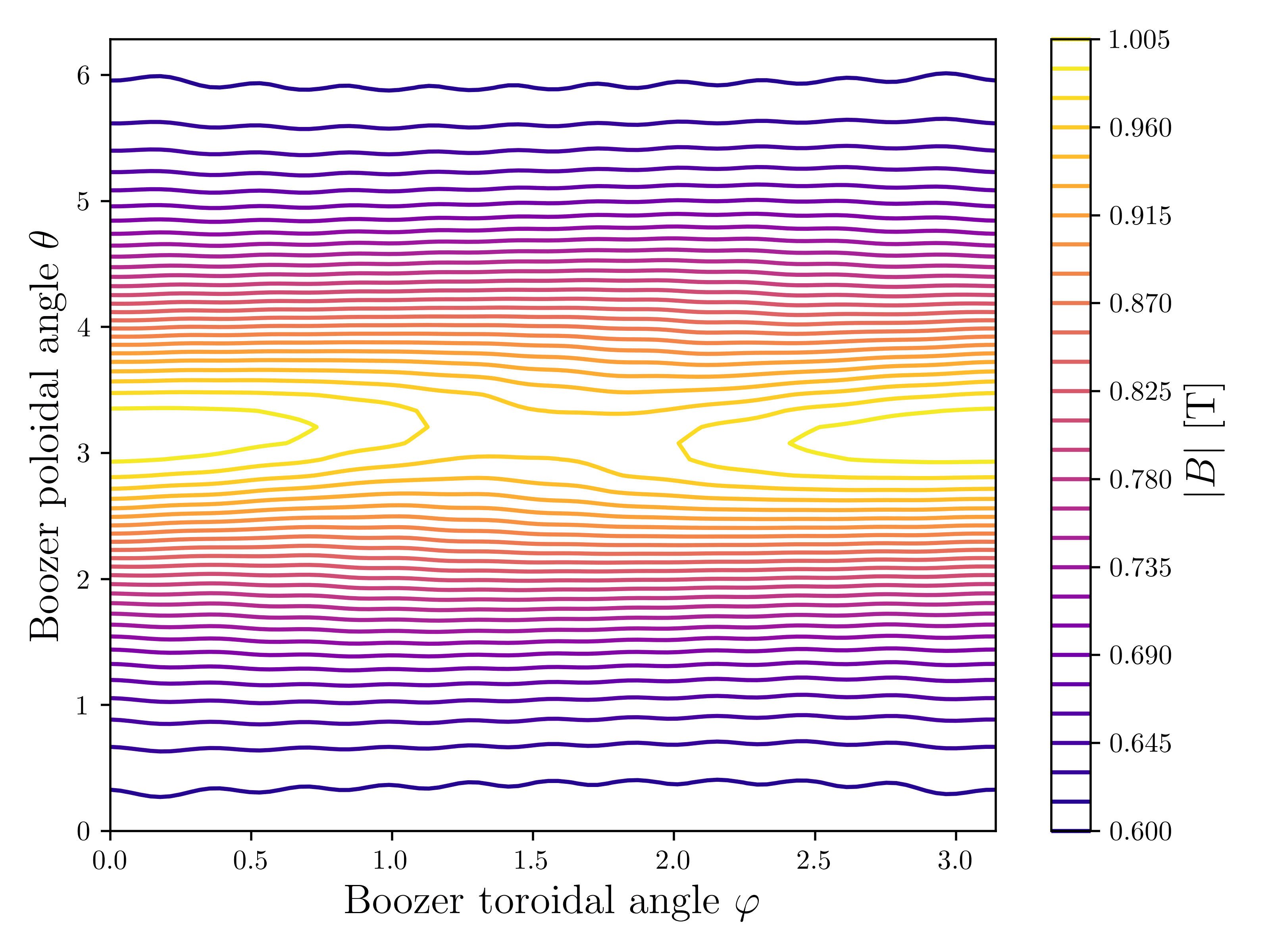}
    \caption{\textbf{Left:} Amplitude of two cost functions during the different optimization stages for the C4\_R19: the squared flux and the departure from quasisymmetry. \textbf{Right:} Boozer plot at the LCFS for the C4\_R19 configuration obtained from a VMEC equilibrium after optimizing a QFM surface.}
    \label{fig:iterations_boozer}
\end{figure}

Finally, in this section, the method for obtaining the best EPOS configuration, C4\_R19, is described. This method is applied to other configurations, such as C4\_R16 and C4\_R18; however, it works with varying degrees of success. This technique is an additional process that follows the Stochastic Stage II, as presented in Figure \ref{fig:iterations_boozer}. The primary purpose of this method is to tailor the geometric characteristics of individual coils that do not present suitable parameters, such as HTS strain, curvature, size, or inter-coil distance. For example, the former C4\_R19 configuration has weave-lane coils that are too big to fit in the vacuum chamber, or the former C4\_R18 has coils 4-8 that possess concave sections. The overall improvement of the squared flux and quasisymmetry is shown in Figure \ref{fig:iterations_boozer} for configuration C4\_R19. The first step corresponds to one optimization where the surface remains unchanged, and circular coils attempt to match it; therefore, QS stays constant, and the squared flux is considerably improved. Then, in the blue section, comes the stochastic single-stage phase, where a trade-off exists between the quality of the quasisymmetry and the accuracy of the coils. After this phase, the equilibria are no longer optimized, and the quasisymmetric error remains constant. Following this, what is referenced here as stage III corresponds to a stochastic stage II, where only the coils are optimized. This yields a visible improvement in the squared flux, going from 1.8e-4 $\text{m}^{2}$ to 4e-6 $\text{m}^{2}$. It should be noted that this step also allows for all the coils to reach the desirable normal curvature everywhere. It requires numerous minor adjustments to the weights, specifically to the coil length penalty and mean-squared curvature. However, other metrics, such as HTS strain and the size of the weave-lanes, are not satisfactory. That is why the final stage IV (in green) corresponds to a phase of adaptive and targeted strategies to correct imperfections. The first step, where a slight increase in the squared flux is visible, refers to the fixing of all the degrees of freedom of the coils, except for those of the weave-lanes coils, together with an increase in the coil length penalty, results in a successful decrease in the size of the weave-lanes to fit in the vacuum chamber. 

\begin{figure} 
    \centering
    \includegraphics[width=\linewidth]{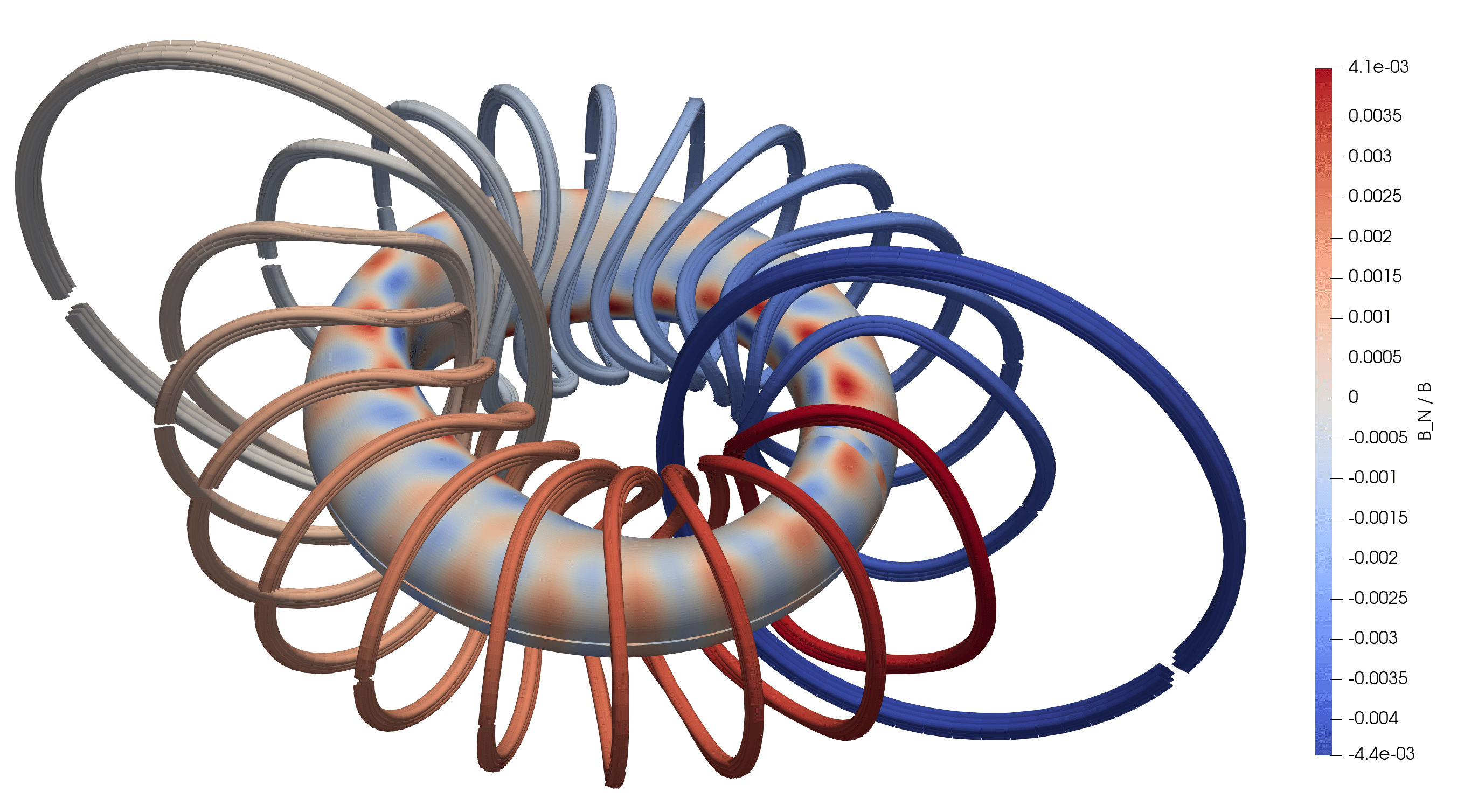}
    \caption{EPOS C4$\_$R18 configuration with coils and LCFS, the heatmap corresponds to the field error $\frac{\langle B\cdot n\rangle}{\langle B \rangle}$. The coils fulfill the engineering requirements and the equilibrium fulfills the physics requirements.}
    \label{fig:C4_R19}
\end{figure}

Moreover, in order to address the loss in squared flux from the previous stage, an additional correction step is performed where the degrees of freedom of the weave-lanes are fixed (preventing them from expanding further) and all the other ones are freed, including the current degrees of freedom of all the coils. This is done as the weave-lanes alone do not allow for considerable improvement of the squared flux (only by a maximal increment of 5e-7 $\text{m}^{2}$) compared to the 10 other coils in the field period. Figure \ref{fig:iterations_boozer} shows the increase in field accuracy yielded by this correction from 5e-6 $\text{m}^{2}$ to 1.8e-7 $\text{m}^{2}$. This enhancement by an order of magnitude is achieved while conserving the normal curvature to desirable degrees, improving the HTS strain levels, and decreasing the current ratio from 4 to 1.8. The last two steps, which slightly deteriorate the squared flux, are taken into account to compensate for the coil-to-coil distance being too small to accommodate the support material. On the right of the optimization procedure in Figure \ref{fig:iterations_boozer} is shown the Boozer plot for the last closed flux surface of the device, displaying well-defined straight lines. The effects of coil ripple are also visible in the small-wavelength perturbations of the lines. The result is one configuration that fulfills all the preliminary requirements for a buildable pair plasma stellarator with HTS; the configuration is shown in Figure \ref{fig:C4_R19}. Planar-like weave-lane coils are a desirable feature, as they simplify the winding procedure, especially for configurations with a current ratio of 1.8, which require only 1.8 times the number of windings of HTS tape as other coils. Furthermore, the weave-lanes are optimized to leave enough space between the inboard side and the weave-lanes as J$\times$B force calculations (see Appendix \ref{sec:forces}) show that if they remain too close to the standard coils, it can create areas where the forces can become problematic for the integrity of the support structure. It should be noted that one advantage for the optimizer of making large weave-lane coils and additionally freeing the coil current degrees of freedom is that it becomes easier to minimize the optimization function by compressing all the other coils together (covering any gap previously due to the high currents of the weave-lanes) and searching for paths to shape the standard coils as if it is a regular stellarator symmetric device, which leads to simpler and repeated coils.


\begin{figure} 
    \centering
    \includegraphics[width = 0.85\linewidth]{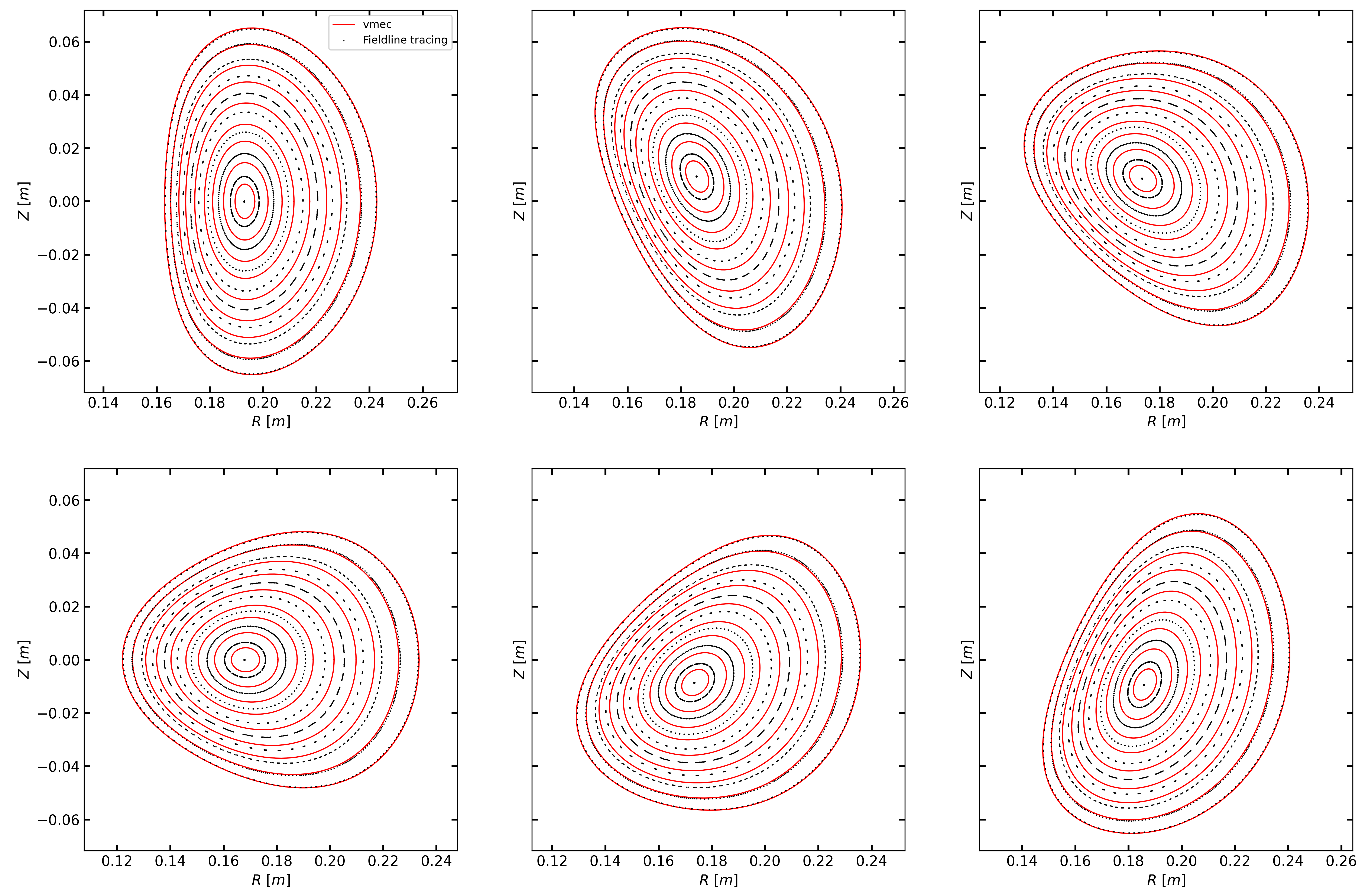}
    \caption{Poincaré cross-sections of the C4\_R19 configuration at four different toroidal positions, comparison between the VMEC solution (red) and field line tracing (black) from the magnetic field generated by coils for different toroidal cross sections, $\phi = [0, \frac{\pi}{6}, \frac{\pi}{3}, \frac{\pi}{2}, \frac{5\pi}{6}, \pi]$. No island chains are visible.}
    \label{fig:poincare}
\end{figure}

The extent of refinement observed in the C4\_R19 configuration is not seen in other configurations where the same technique is applied. However, other geometric issues could be resolved as mentioned previously. This shows that targeting localized issues of the configuration works as a straightforward method to exit local minima and access other. Fixing specific degrees of freedom acts as a restriction of the parameter space, which facilitates the search for the optimizer. Additionally, it should be noted that the primary factor driving the continuous adjustment of the weights and necessitating ongoing refinement is the curvature of the coils.
In most cases, further improvement of the squared flux requires a trade-off with the concavity of the coils. However, this is not the case for the C4\_R19 configuration. One hypothesis explaining the available minimum is that the equilibrium LCFS is convex and does not exhibit significant shaping, allowing for simple coils to be found. The poincaré cross-sections of this configuration are shown in \ref{fig:poincare}. No visible island chains are present; the flux surfaces do not possess considerable shaping, as the average iota is 0.06.

\begin{figure}
    \centering
    \includegraphics[width=0.7\linewidth]{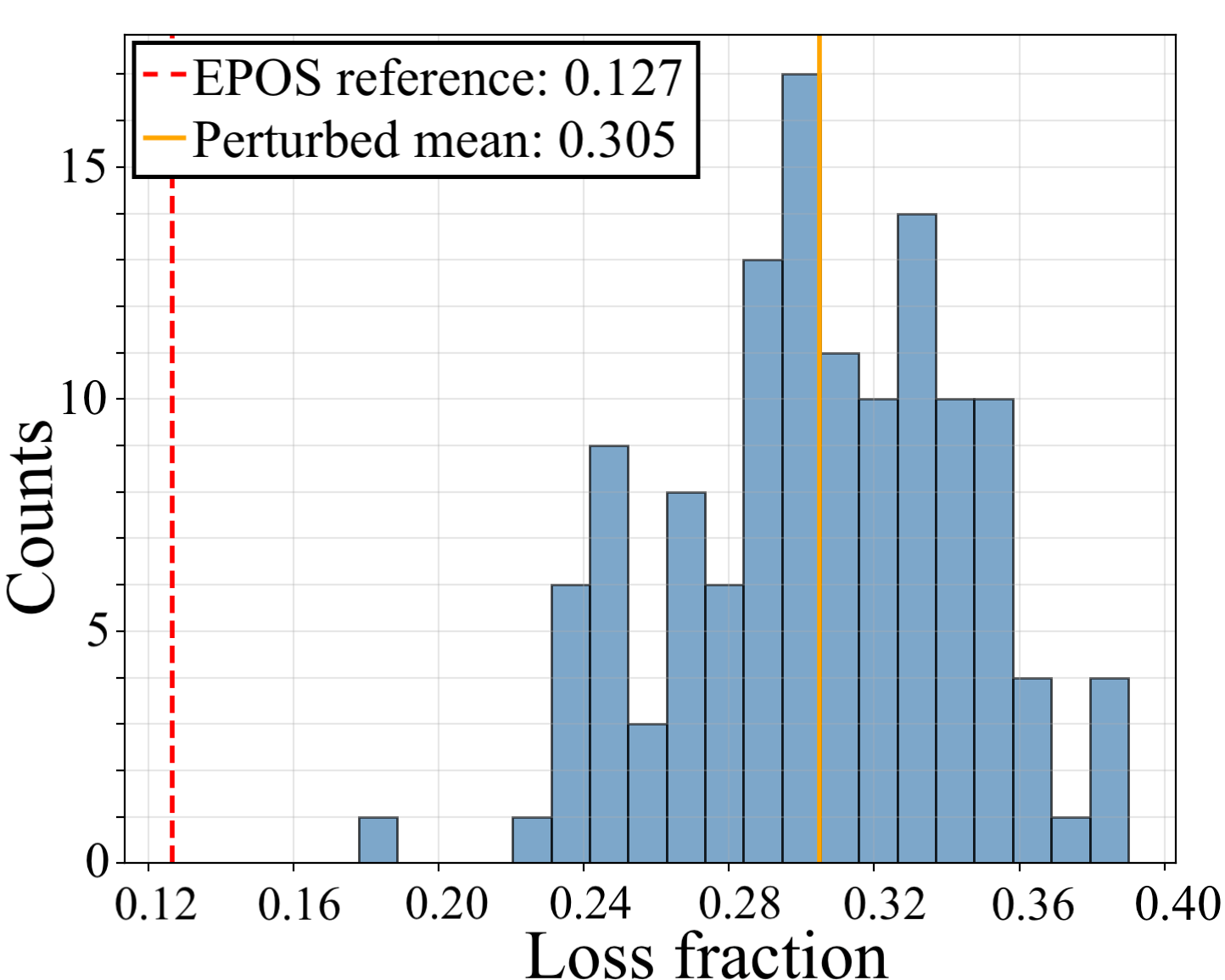}
    \caption{Loss fraction distribution for 128 perturbed EPOS C4\_R19 configurations. Each count represents one EPOS stellarator where the coils have been randomly perturbed with 1 mm deviations. A thousand particles in the simulation are followed for 0.5 seconds with 1 eV each starting at the radial coordinate of s=0.9.}
    \label{fig:particle_robustness}
\end{figure}

An additional check of how manufacturing and assembly deviations directly affect particle confinement is presented in Figure \ref{fig:particle_robustness}. The distribution of the loss fraction of particles is shown after following 1000 trajectories for 0.5 seconds with SIMPLE. The particles carry an energy of 1 eV and are initialized on the outboard side of the stellarator at a radial coordinate of s=0.9. The distribution is compared to the unperturbed C4\_R19 configuration which loses 12.7\% of the particles. Perturbing the coils with 1 mm deviations deteriorates the confinement and results in an average loss of 30\% of electrons and positrons reaching in some cases 38\%. These results do not account for two experimental factors that work in favor of better confinement: the velocity distribution used in these simulations is random, however the positron beam feeding the experiment is expected to have on average a greater parallel velocity than perpendicular. Then, the simulations do not account for cyclotron cooling that radiates a considerable amount of the perpendicular energy. Given the dimensions of the stellarator, having 30\% particle loss after 0.5 seconds with coil perturbation amplitudes of 1 mm, with random perpendicular velocities and no cyclotron cooling, is a promising result.

Further post optimization studies regarding the quality of the C4\_R19 configuration are also performed through large-scale scans where 12600 different EPOS stellarators are considered (see Annex \ref{sec:epos_scans}). These scans enable us to compare, at least locally, the results obtained so far. The primary coil characteristics of C4\_R19 seem to outperform all the configurations from the scans. Although the quasisymmetric error of this configuration is higher than most configurations found in the scans, what makes it stand out particularly is the field accuracy that can be achieved with relatively simple coils.

\section{Conclusion}
\label{sec:conclusion}

In this paper, the procedure to find suitable candidates for the EPOS stellarator is presented. With the ultimate goal of confining pair plasmas, the final configuration is able to fulfill a series of physics and engineering constraints that restrict the parameter space of possible buildable stellarators. A determining factor for EPOS is the size requirement needed to achieve densities high enough to observe collective behavior.
Concretely, this is expressed in the search for configurations that possess low iota for coils with low HTS strain, no concavity, weave-lanes with stray fields for the injection of particles, sufficient coil-to-coil distance to accommodate material between them, and improved robustness to manufacturing errors of the coils. These conditions led to the development of a minimization scheme that includes stochastic optimization, as well as single-stage methods and stage II techniques with finite build coils, HTS strain, and a subsequent restriction of the parameter space. 

To assess different configurations, a scan is performed at the major radius and in the current ratio between the standard coils and the weave-lanes. Although the major radii vary only between 16 and 19 cm and the current ratios between 3 and 4, considerable differences are visible among all the configurations, providing an indication of the breadth of the optimization space for 3D configurations. Single particle guiding center orbit simulations showed that confinement times of about 2 seconds with losses inferior to 10 \% are possible when starting the particles at the magnetic axis. Moreover, a posteriori robustness verification for the configurations not only showed that some are tolerant to perturbations exceeding 1 mm but also revealed an indication of intrinsic robustness in severalconfigurations. Finally, all the candidates achieved below threshold values for the HTS strain, and all those with a current ratio of 4 found convex coils. Specifically, EPOS C4\_R9 displayed good characteristics in all the metrics and will be studied further. This study presents a possible approach to optimizing the design of a stellarator, enabling its complete construction. Following this design stage, a subsequent engineering phase is being conducted to assess the feasibility of the configuration \citep{gil_manufacturing_2025, huslage_non-planar_2025}. This includes analysis of the stresses on the coils, the magnetic field permeability of the materials used in construction, and even positron injection under realistic conditions to enable a comprehensive design of the coils.

Looking towards future devices, the integration of engineering proxies is highly valuable as it can guide optimization, taking into account, for example, von Mises/Tresca stresses instead of solely forces, field errors from magnetic permeability of materials, and current distribution in the winding packs, etc. 

Finally, if EPOS achieves its density targets, due to average ``bad curvature", there is a possibility that the electrostatic version of interchange modes can still arise. This, however, would not be catastrophic, quite the opposite in fact, as it would be the first time MHD instabilities were observed in a matter-antimatter plasma.

\newpage
\appendix

\section{Number of Coils}
\label{sec:num_coils}
To determine the number of coils in the configurations studied above, a study is conducted on the quality of quasisymmetry as a function of the number of coils per half-field period and the distance between coils and the last closed flux surface. 

\begin{figure}
    \centering
    \includegraphics[width=\linewidth]{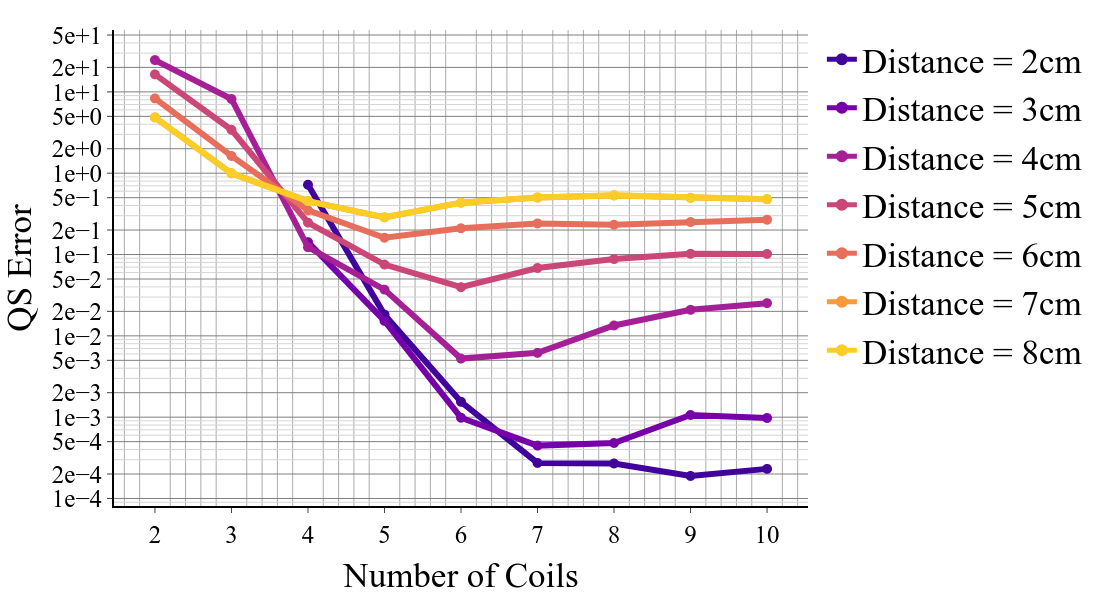}
    \caption{Quasisymmetric error as a function of the number of coils and of the coil to surface distance.}
    \label{fig:qsvsncoils}
\end{figure}

The REGCOIL optimization software \citep{regcoil} is employed. REGCOIL solves a convex optimization problem, where it tries to find the current potential that best reproduces the target plasma surface on a predetermined coil winding surface. One of the advantages of this procedure is that it has only one global minimum, thereby allowing for accurate comparisons between different coil-to-surface distances, for example. Once the current potential is found, the software is then able to ``cut coils" from the winding surface, and here is where the number of coils is set by the user. Then, once a configuration with coils is found, a check of the quality of the quasisymmetry is performed using the previously mentioned QFM method. Figure \ref{fig:qsvsncoils} summarizes the findings for this EPOS configuration.

As expected, the quasisymmetric error drops as the number of coils increases. This can be explained by the reduction of coil ripple due to the increasing coverage of the coils. However, after seven coils, it is visible that there is a saturation in the quality of quasisymmetry, which is observed across all coil-to-surface distances. Moreover, if the 2 cm coil-to-surface distance is not taken into account, saturation is even achieved at six coils per half-field period. Since six coils per half-field period with finite dimensions are too cumbersome for the device, especially at 16 cm in major radius, it is decided to set the number of coils at 11 coils per field period, including weave-lane coils.

\section{Coil Forces}
\label{sec:forces}

\begin{figure} 
    \centering
    \includegraphics[width=0.35\linewidth]{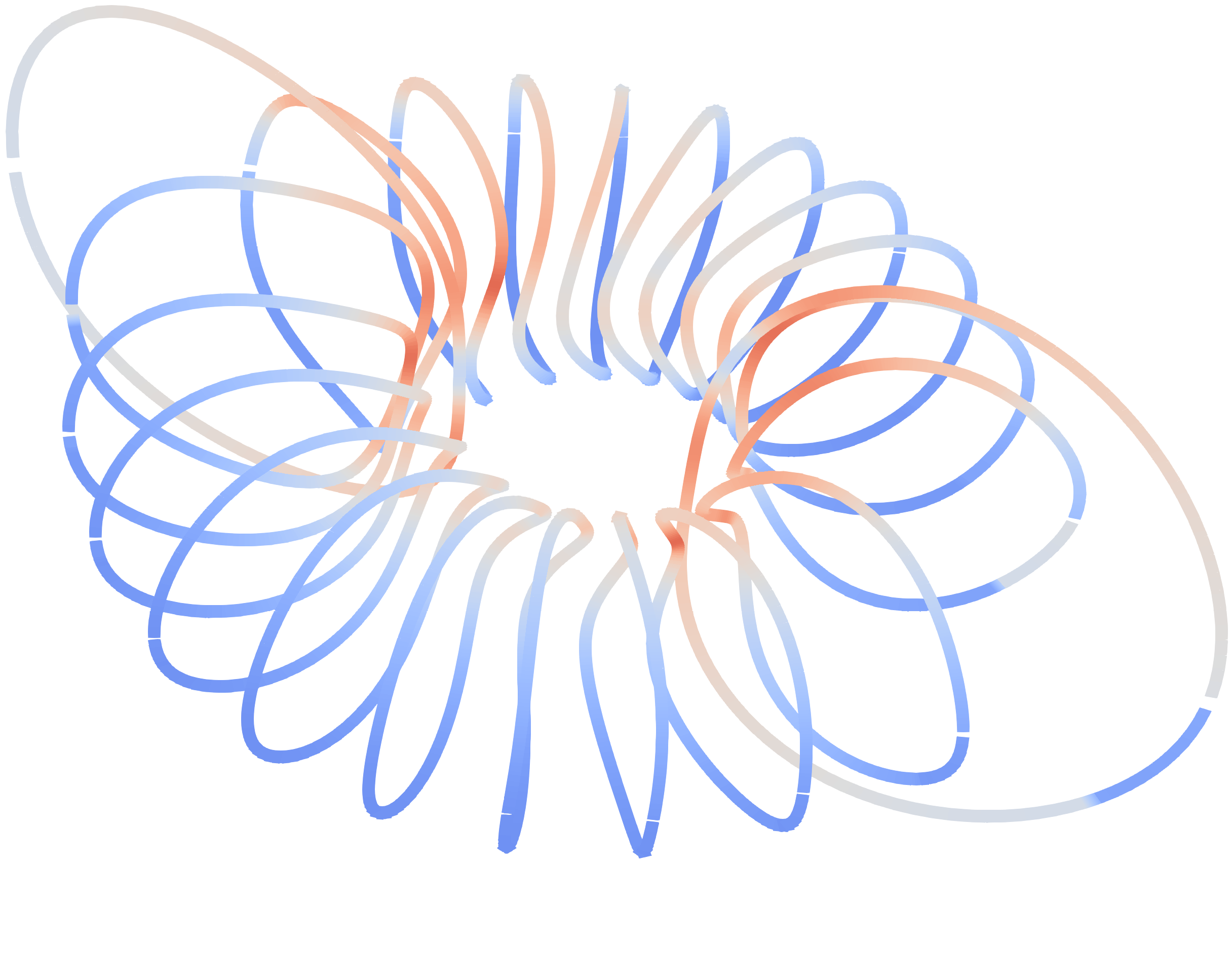}
    \includegraphics[width=0.45\linewidth]{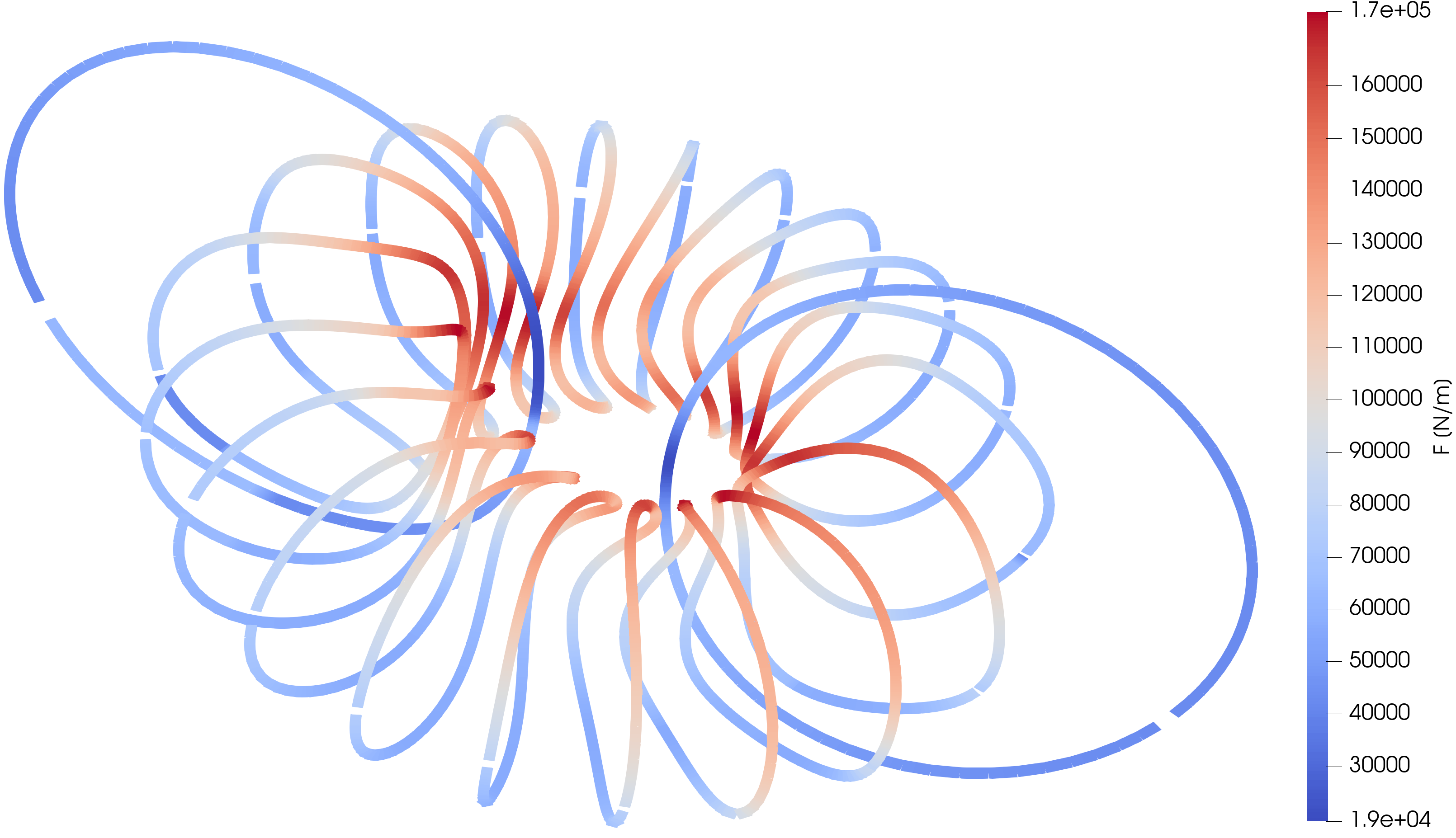}
    \caption{Simulated coil forces showed on the coil filaments expressed in N$\cdot$m on two EPOS candidates: C3\_R16 (\textbf{left}) and C4\_R19 (\textbf{right}), spatial dimensions are to scale, apparent discontinuities in the left plot are due to applying a common scale for both stellarators.}
    \label{fig:epos_forces}
\end{figure}

Given the small dimensions and high fields of the EPOS device, mechanical stresses due to J$\times$B forces are checked in this section. Using the force calculation framework described in \citep{Landreman_2025} and implemented in SIMSOPT \citep{hurwitz2024electromagneticcoiloptimizationreduced}, a 3D force vector in each of the quadrature points of each coil, a 3D force vector is calculated in N$\cdot$m. This vector includes contributions from forces due to the interaction between the current in a given coil and its own generated field (self-field), as well as the external field induced by the remaining coils. These calculations are obtained by assuming a coil ``wire" with a finite rectangular cross section of 6 mm × 12 mm, which is close to the expected experimental values of EPOS. In Figure \ref{fig:epos_forces}, the amplitude of the forces for two different candidates, C3\_R16 (\textbf{left}) and C4\_R19, the latter being the best candidate so far, is displayed. 

These two configurations show possible differences in feasibility between two design choices: C3\_R16 has a smaller major radius and coils (when compared to C4\_R19), which can lead to smaller coil-to-coil distances and higher forces. However, because of the 2 T restriction on the axis, this also means that larger coils require larger currents. Combined, this leads to comparable amplitudes between the two configurations, with the maximum value for C3\_R16 being 1.5e5 N$\cdot$m and for C4\_R19 1.7e5 N$\cdot$m. The overall higher force values for the C4\_R19 are observed in the inboard side, which is expected since the coil-to-coil distance decreases in that region. One additional feature of the larger configuration is the spacing between the inboard side of the standard coils and the weave-lane coils. Although most coils experience increased stresses, the weave-lane coils exhibit considerably reduced forces compared to their smaller counterparts. This is mainly due to two factors: first, as mentioned previously, C4\_R19 only possesses a factor of 1.8 between the currents of the standard coils and the WL coils, and unlike C3\_R16 on the inboard side, the WL coils are further apart from the standard coils.

\section{Scans of the configuration space}
\label{sec:epos_scans}
Following the 8 EPOS configuration study presented in previous sections, where the configurations are extensively optimized to present viable candidates, large-scale scans allow us to observe where these candidates exist in the N-dimensional stellarator configuration space. The principle of the scans presented here is to target large numbers of possible EPOS configurations without requiring that they be buildable, in order to sample from different regions in the parameter space, possibly providing unexplored better configurations. Note that, unlike the optimization explained in this work, the scans are performed following a traditional stage-I/stage-II method.

\begin{figure} 
    \centering
    \includegraphics[width=0.47\linewidth]{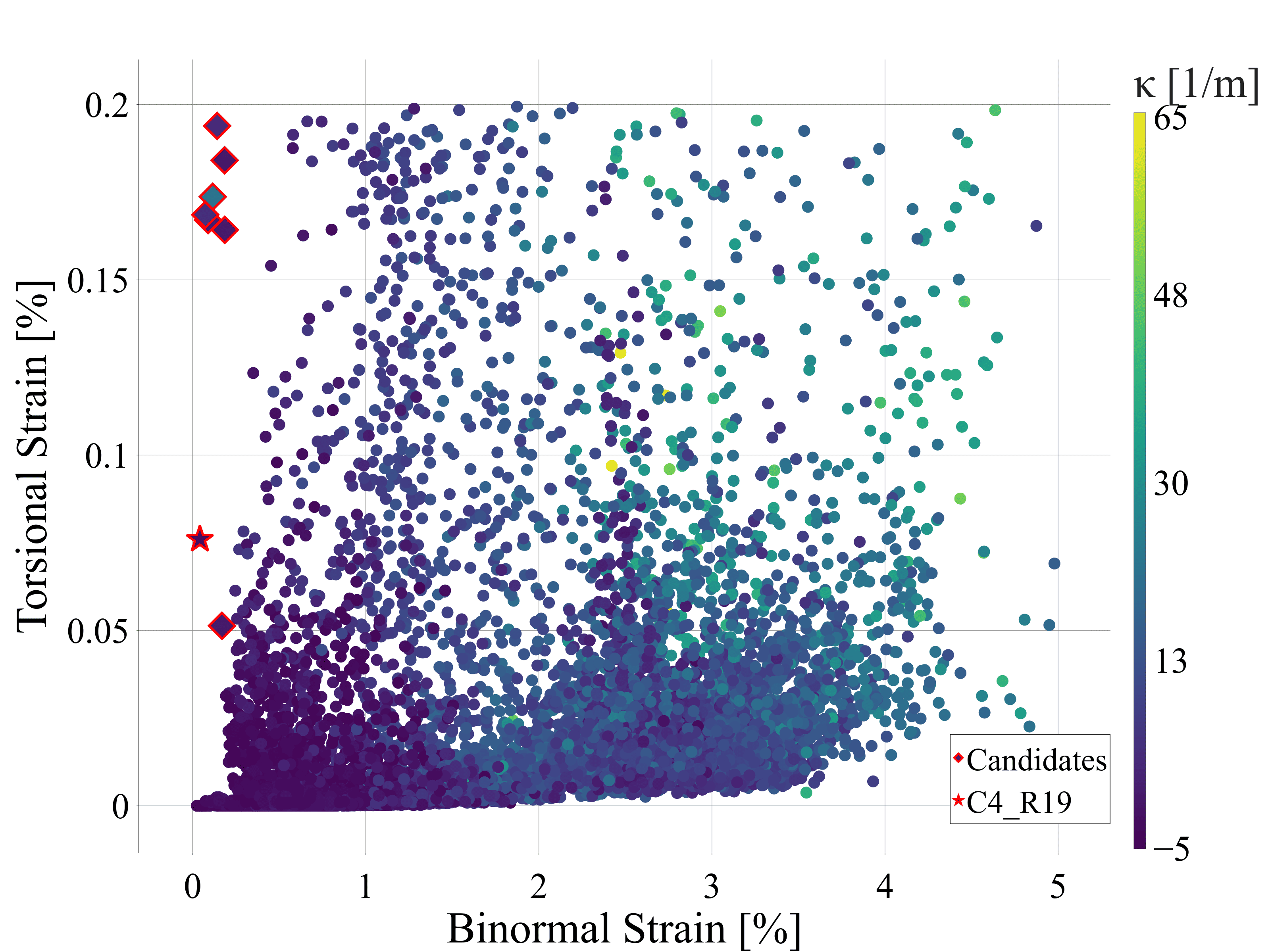}
    \includegraphics[width=0.47\linewidth]{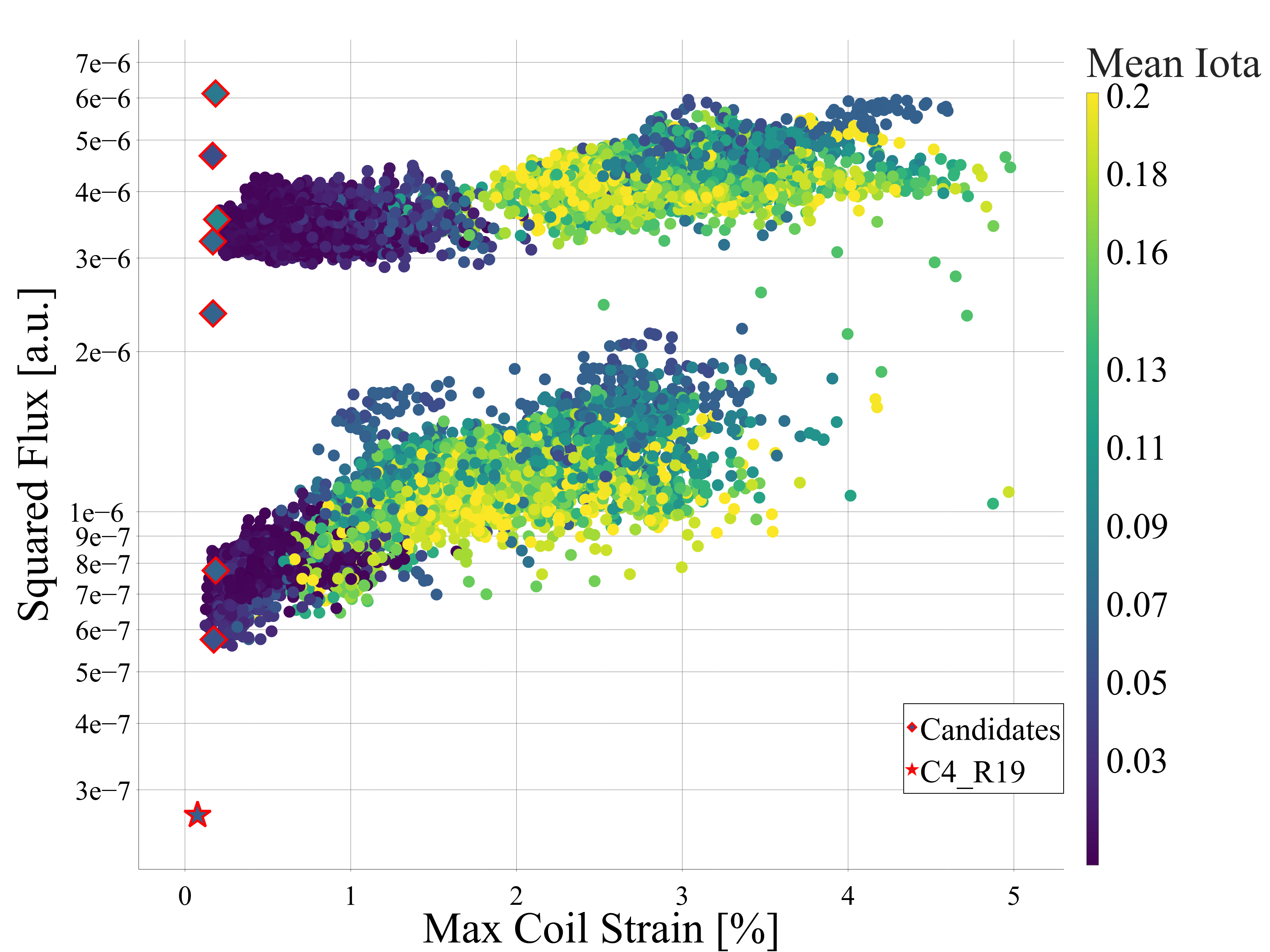}
    \caption{\textbf{Left:} Scatter plot showing the torsional strain versus the binormal strain, where the intensity shows the convexity of the coils of the EPOS large-scale scans. \textbf{Right:} Squared flux versus the maximum measured coil strain of a configuration defined as the maximum of the binormal and the torsional strain together with the mean iota of each configuration as a heatmap.}
    \label{fig:epos_scans}
\end{figure}

 Main target parameters for the EPOS equilibria are: major radii between 16 and 19 cm, iota between 0.05 and 0.20, aspect ratio between 3 and 6, the magnetic well is also varied by varying the magnitude of its respective hyperparameter, as for the coils the ratio between the weave-lane coils and the standard coils is scanned between 2 and 4, and the coil length is scanned through its hyperparameter variation. The total number of different coilsets found is 12600, with effectively 2100 different EPOS equilibria, and six different coilsets for each equilibrium. Note that these numbers are still low compared to the uncountable measure of the parameter space. 
Nonetheless, the information provided by these scans allows us to place the configurations found in this work at least locally. As observed in Figure \ref{fig:epos_scans}, the left plot shows the three most important quantities regarding the coils: torsional versus binormal strain and coil convexity $\kappa$. With the red outlines, the data points referring to the fully optimized configurations are plotted. Compared to the 12600 data points they display optimal strains and are located in a region unreached by the scans at very low binormal strains ( $\epsilon \leq$ 0.2 \% ) and convex coils ($\kappa$ $\leq$ 0). Moreover, on the right plot, where the quadratic flux is plotted against the maximum coil strain, and considering the rotational transform of the configurations, these configurations appear again at low strain, with similar field accuracy, but at a higher iota. Note that the C4\_R19 configuration stands out in all of the parameters shown here.

\section{Parameters of the C4\_R19 configuration}

\begin{table} 
\centering
\begin{tabular}{lllllc}
	\hline
	\multicolumn{1}{l}{Parameters}                       &  &  & & & \multicolumn{1}{c}{C4\_R19} \\ \hline
	Max $ B\cdot n /B$ &  &  & & &     4.1e-3  \\ \hline
	$\langle B\cdot n\rangle / \langle B \rangle$ &  &  & & &     9.6e-4                      \\ \hline
	QS error [a.u]                             &  &  & & &        1.1e-3                    \\ \hline
	QS at $\sqrt{3}\sigma$ = 1mm  [a.u.]                      &  &  &  & &     4e-3                   \\ \hline
	Minor Radius [cm]  &  &  &  & &  5.12 \\ \hline
	Major Radius [cm] &  &  &  & &  19 \\ \hline
	Aspect Ratio &  &  &  & &  3.71 \\ \hline
	Mean Rotational Transform $\iota$ &  &  &  & & 0.064 \\ \hline 
	Well [\%] &  &  &  & &  -21.3 \\ \hline
	Required Positrons at T$_{e}$ = 1eV &  &  &  & &  $\sim$ 1e10\\ \hline
	a / $\lambda_{D}$ at T$_{e}$ = 1eV &  &  &  &  &9.73 \\ \hline
	Equilibrium Volume [L] &  &  &  & &  9.87 \\ \hline
	Avg Coil Length (Std/WL) [cm]                                &  &  & & &        86 / 154                    \\ \hline
	Total Coil Length [m]                                &  &  & & &        9.53                    \\ \hline
	Min Coil-to-Plasma Distance [cm] &  &  &  & &  $\sim$ 3.7* \\ \hline
	Min. Coil-to-Coil Distance [cm] &  &  &  & &  $\sim$ 1.5* \\ \hline
	Max. Force on Conductor [N/m] &  &  &  & &  1.7e5 \\ \hline
	Currents (Std/WL) [kA] &  &  &  & &  77.7 / 142.7\\ \hline
	Max Binormal Strain [\%] &  &  &  & & 0.041 \\ \hline
	Max Torsional Strain [\%] &  &  &  & & 0.076 \\ \hline
	Max Concavity $\kappa$ [m$^{-1}$] &  &  &  & &  -2.57 \\ 
	\hline
	Max. Curvature [m$^{-1}$] & & & & & 18.9 \\
	\hline
	Max. MSC Curvature [m$^{-1}$] & & & & & 103.57 \\
	\hline
\end{tabular}
\caption{Main parameters of the C4\_R19 configuration of EPOS, standing as the best candidate so far. *Note that coil-to-surface and coil-to-coil distance values consider already SIMSOPT finite-build modifications; however, these are still subject to variations depending on the final engineering design that accounts for finite CAD dimensions.}
\label{tab:epos_table}
\end{table}

\newpage
\section{Overview of the EPOS candidates configurations}
\label{sec:spider_plots}
\begin{figure} 
	\centering
	
	
	\begin{subfigure}[t]{0.45\textwidth}
		\includegraphics[width=\textwidth]{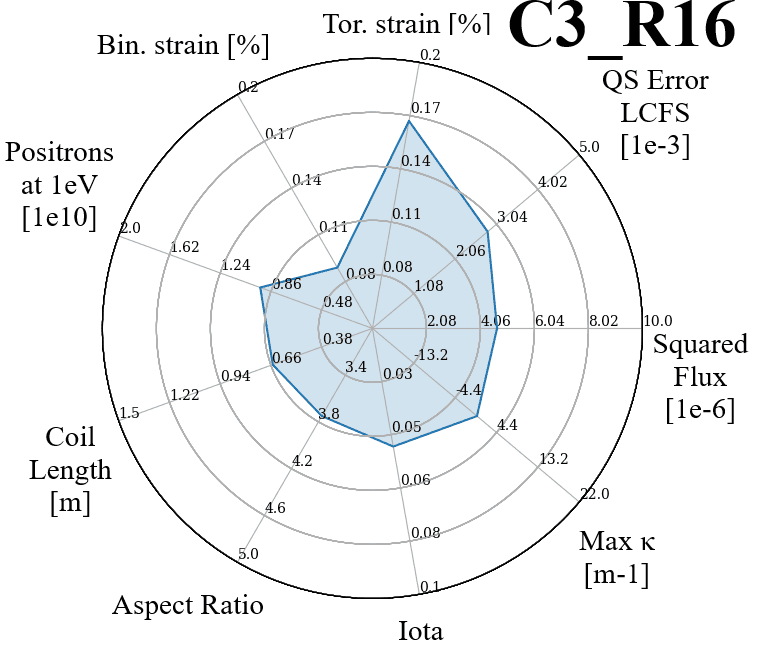}
		\caption{}
		\label{fig:c3_r16}
	\end{subfigure}
	\hspace{0.5cm}
	\begin{subfigure}[t]{0.45\textwidth}
		\includegraphics[width=\textwidth]{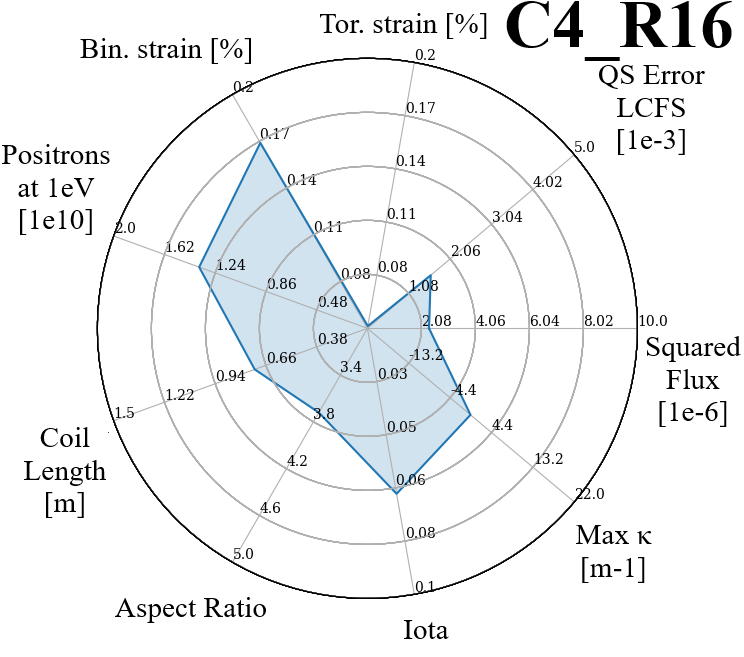}
		\caption{}
		\label{fig:c4_r16}
	\end{subfigure}
	
	\par\bigskip
	
	\begin{subfigure}[t]{0.45\textwidth}
		\includegraphics[width=\textwidth]{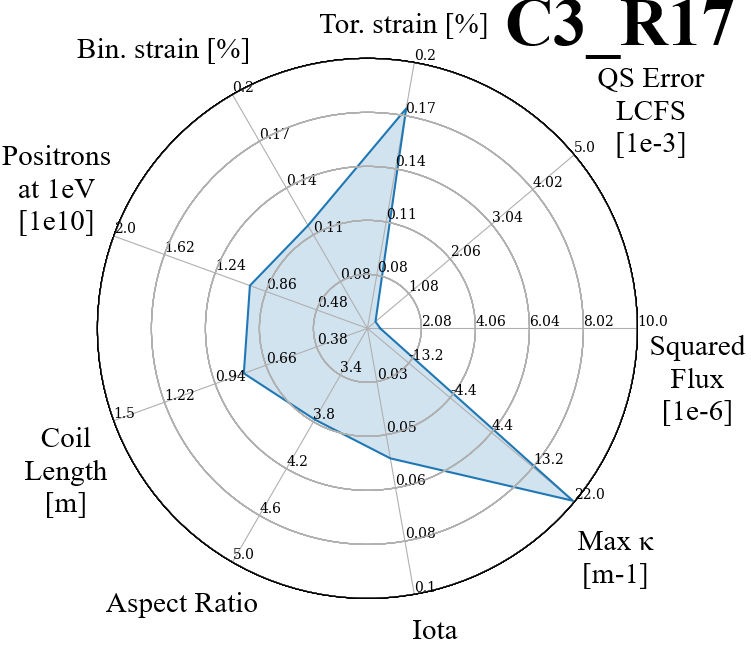}
		\caption{}
		\label{fig:c3_r17}
	\end{subfigure}
	\hspace{0.5cm}
	\begin{subfigure}[t]{0.45\textwidth}
		\includegraphics[width=\textwidth]{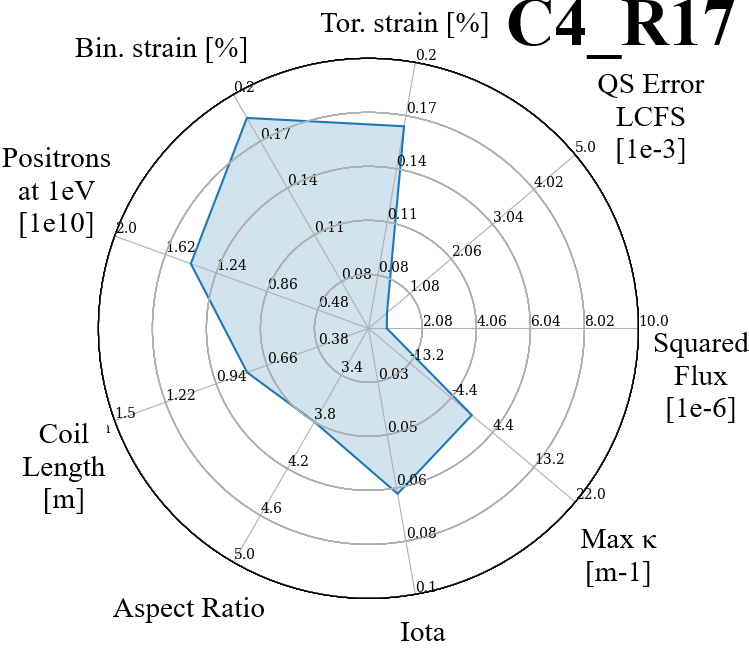}
		\caption{}
		\label{fig:c4_r17}
	\end{subfigure}
	
	\caption{Spider plots summarizing the relevant properties for the EPOS stellarator candidates with major radii 16 and 17 cm. For all the properties except for iota, the blue area should ideally be minimized.}
	\label{fig:qh_coils}
\end{figure}

\begin{figure} 
	\centering
	
	
	\begin{subfigure}[t]{0.45\textwidth}
		\includegraphics[width=\textwidth]{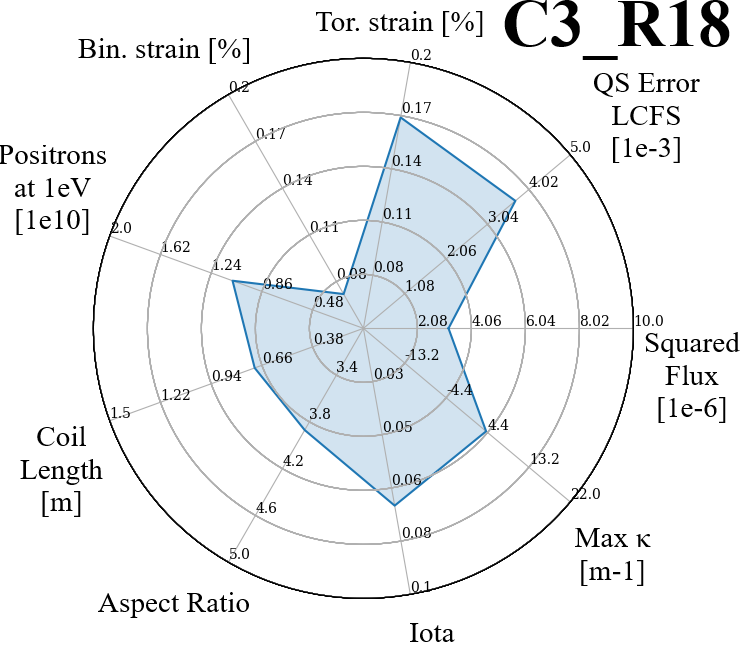}
		\caption{}
		\label{fig:c3_r18}
	\end{subfigure}
	\hspace{0.5cm}
	\begin{subfigure}[t]{0.45\textwidth}
		\includegraphics[width=\textwidth]{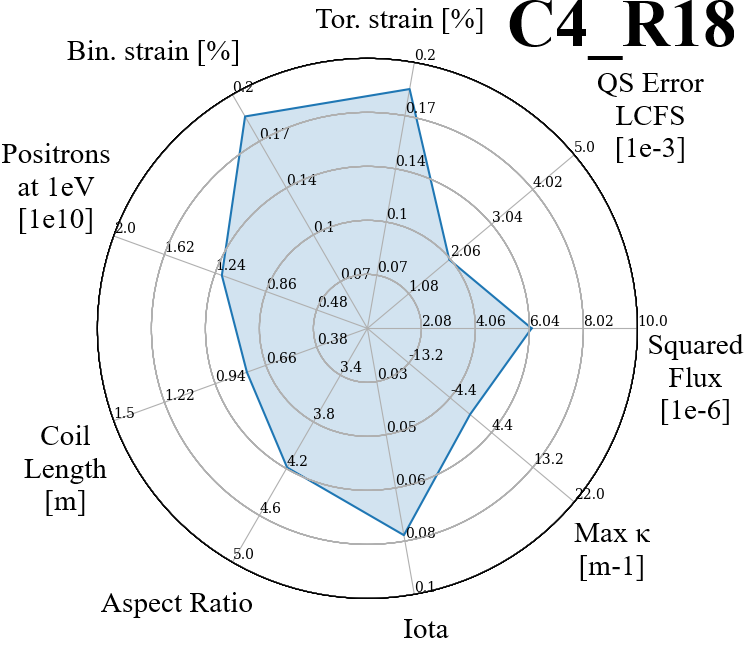}
		\caption{}
		\label{fig:c4_r18}
	\end{subfigure}
	
	\par\bigskip
	
	\begin{subfigure}[t]{0.45\textwidth}
		\includegraphics[width=\textwidth]{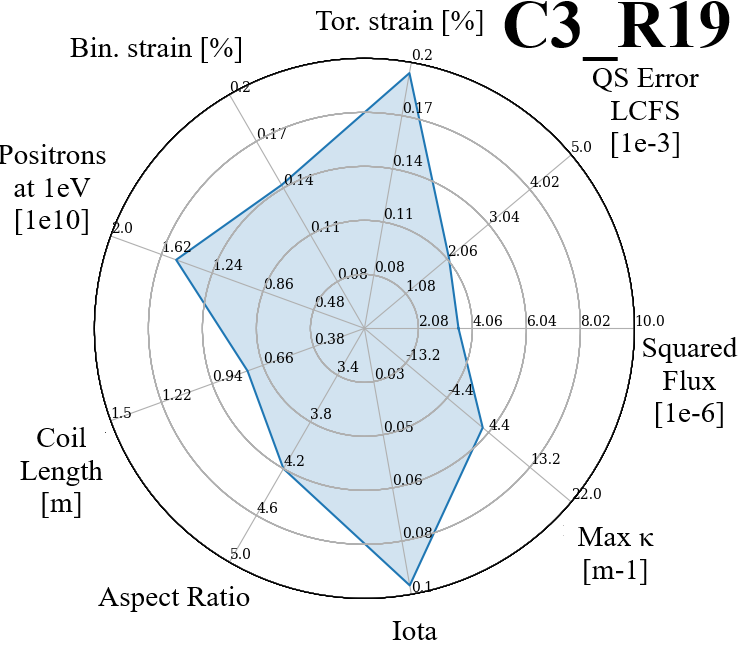}
		\caption{}
		\label{fig:c3_r19}
	\end{subfigure}
	\hspace{0.5cm}
	\begin{subfigure}[t]{0.45\textwidth}
		\includegraphics[width=\textwidth]{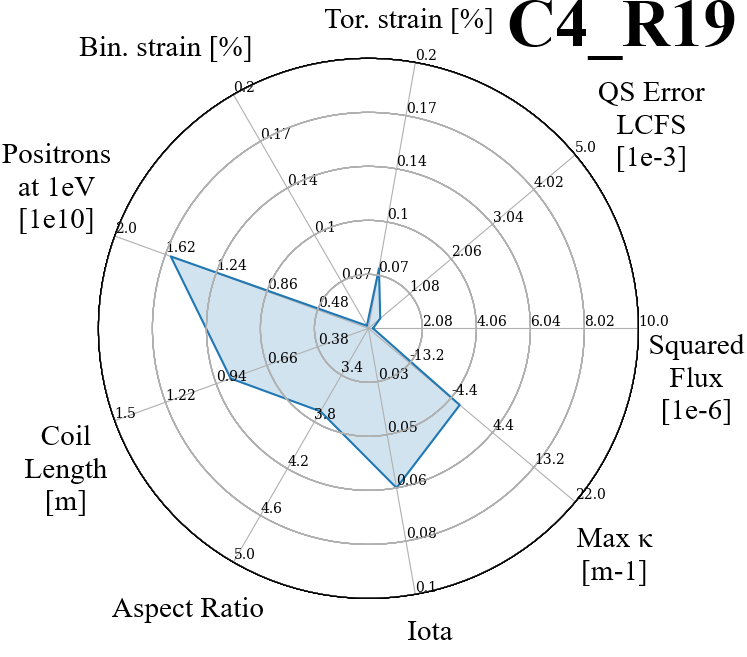}
		\caption{}
		\label{fig:c4_r19}
	\end{subfigure}
	
	\caption{Spider plots summarizing the relevant properties for the EPOS stellarator candidates with major radii 18 and 19 cm. For all the properties except for iota, the blue area should ideally be minimized.}
	\label{fig:spider_plots}
\end{figure}

\newpage
\bibliographystyle{jpp}

\bibliography{jpp-instructions}

\end{document}